\documentclass[prb,twocolumn,aps,floatfix]{revtex4-2}
\usepackage[utf8]{inputenc}
\usepackage[caption=false]{subfig}
\usepackage{graphicx,xcolor}
\usepackage{multirow}
\usepackage{amsmath}
\usepackage{amssymb}
\usepackage{braket}
\usepackage{soul}
\usepackage[normalem]{ulem}
\usepackage{dsfont}
\usepackage{hyperref}

\newcommand{\Tr}{\textrm{Tr}}

\begin{document}

\title{Symmetry Resolved Entanglement Entropy in a Non-Abelian Fractional Quantum Hall State}

\author{Mark J.\ Arildsen}
\affiliation{SISSA --- International School for Advanced Studies and INFN, via Bonomea 265, 34136 Trieste, Italy}

\author{Valentin Cr\'{e}pel}
\affiliation{Center for Computational Quantum Physics, Flatiron Institute, 162 5th Avenue, New York, NY 10010, USA}
\affiliation{Department of Physics, University of Toronto, 60 St.~George Street, Toronto, ON, M5S 1A7 Canada}

\author{Nicolas Regnault}
\affiliation{Center for Computational Quantum Physics, Flatiron Institute, 162 5th Avenue, New York, NY 10010, USA}
\affiliation{Laboratoire de Physique de l'Ecole normale sup\'{e}rieure, ENS, Universit\'{e} PSL, CNRS, Sorbonne Universit\'{e}, Universit\'{e} Paris-Diderot, Sorbonne Paris Cit\'{e}, 75005 Paris, France}
\affiliation{Department of Physics, Princeton University, Princeton, New Jersey 08544, USA}

\author{Benoit Estienne}
\affiliation{Sorbonne Universit\'e, CNRS, Laboratoire de Physique Th\'eorique et Hautes Energies, LPTHE, F-75005 Paris, France}

\begin{abstract}
Symmetry-resolved entanglement entropy provides a powerful framework for probing the internal structure of quantum many-body states by decomposing entanglement into contributions from distinct symmetry sectors. In this work, we apply matrix product state techniques to study the bosonic, non-Abelian Moore-Read quantum Hall state, enabling precise numerical evaluation of both the full counting statistics and symmetry-resolved entanglement entropies. Our results reveal an approximate equipartition of entanglement among symmetry sectors, consistent with theoretical expectations and subject to finite-size corrections. The results also show that these expectations for symmetry-resolved entanglement entropy remain valid in the case of a non-Abelian state where the topological sectors cannot be distinguished by the Abelian $\mathrm{U}(1)$ symmetry alone, and where neutral and charged modes possess distinct velocities. We additionally perform a detailed comparison of the entanglement spectrum with predictions from the Li-Haldane conjecture, finding remarkable agreement, and enabling a more precise understanding of the effects of the distinct neutral and charged velocities. This not only provides a stringent test of the conjecture but also highlights its explanatory power in understanding the origin and structure of finite-size effects across different symmetry sectors.
\end{abstract}

\maketitle

\section{Introduction}

Quantum entanglement is an essential aspect of quantum systems that is a very useful theoretical instrument for understanding physical phenomena possessing quantum correlations, both in condensed matter and high energy physics~\cite{Amico2008,Calabrese2009,Laflorencie2016,Rangamani2017}. For instance, the entanglement entropy (EE) of gapped phases of matter follows an area law analogous to that found in black holes~\cite{Srednicki1993,Eisert2010}. One-dimensional critical systems, on the other hand, satisfy a logarithmic rule proportional to the central charge~\cite{Holzhey1994,Vidal2003,Calabrese2004}. Entanglement measures have also been used for the quantum Hall effect (QHE), where they can probe intrinsic topological order~\cite{KitaevPreskill2006,Levin2006} or identify gapless edge modes at boundaries~\cite{Estienne2020,EstienneOblak2021} and interfaces of distinct fractional quantum Hall states~\cite{Crepel2019a,Crepel2019b,Crepel2019c}. 

Entanglement also has a fascinating interplay with the symmetries present in quantum states. This is the preserve of \textit{symmetry-resolved entanglement}. The effect on entanglement of fluctuations of the local charge of an internal symmetry have long been studied~\cite{Song2010,Song2012,Petrescu2014}. The random variable describing these fluctuations is known as the \textit{full counting statistics} (FCS)~\cite{Levitov1993,Levitov1996}. The FCS can be used to find the Luttinger parameter of 1D systems~\cite{Estienne2020,Rachel2012,Hackenbroich2021}, keep track of massless Dirac fermions in 2D~\cite{Crepel2021}, and measure the long-wavelength limit of the structure factor of gapped 2D liquids~\cite{Estienne2022}. Further, cold atom and ion trap experiments have recently demonstrated that analysis of entanglement in different symmetry sectors can illuminate properties of many-body quantum systems~\cite{Lukin2019,Azses2020a,Neven2021,Vitale2022}. 
An understanding of the decomposition of entanglement in symmetry sectors of fixed charge can be provided by symmetry-resolved entanglement measures~\cite{Laflorencie2014,Wiseman2003,Goldstein2018,Xavier2018,Calabrese2021}, which provide a more detailed understanding than total entanglement measures not explicitly sensitive to symmetry. Symmetry-resolved entanglement measures, and in particular symmetry-resolved entanglement entropies (SREE) and their connections to charge fluctuations, have by now been extensively studied in a wide variety of contexts: critical~\cite{Goldstein2018,Xavier2018,Feldman2019,Cornfeld2018,Murciano2021,Capizzi2020,Bonsignori2021,EstienneIkhlef2021,Chen2021,Capizzi2021,Parez2021a,Jones2022,Ares2022b,DiGiulio2023} or gapped~\cite{Barghathi2019,Calabrese2020} 1D systems, systems of free particles~\cite{Crepel2021,Barghathi2018,Tan2020,Bonsignori2019,Fraenkel2020,Murciano2020,MurcianoRuggieroCalabrese2020,KieferEmmanouilidis2020,Horvath2021,Parez2021b,Fraenkel2021,Chen2022,Ares2022a,Foligno2023,Capizzi2023b}, integrable models~\cite{MurcianoDiGiulioCalabrese2020,Horvath2020,Horvath2022,CapizziHorvathCalabrese2022,Piroli2022,Capizzi2022a,Capizzi2022b,Capizzi2023a,CastroAlvaredo2023}, holographic and gravitational systems~\cite{Zhao2021,Weisenberger2021,Zhao2022,Baiguera2022}, and topological phases of matter~\cite{Cornfeld2019,Monkman2020,Azses2020a,Azses2020b,Azses2021,Oblak2022,Milekhin2023}. 

It has been observed that the symmetry-resolved EE is usually distributed evenly among symmetry sectors for typical charge fluctuations, a phenomenon referred to as equipartition of entanglement~\cite{Laflorencie2014,Xavier2018}. 
This behavior has been confirmed in many 1D systems, e.g.~in Refs.~\cite{Xavier2018,MurcianoDiGiulioCalabrese2020,Turkeshi2020}, including non-Abelian Wess-Zumino-Witten models~\cite{Calabrese2021}, and in free systems in 2D~\cite{MurcianoRuggieroCalabrese2020}. 
Entanglement equipartition has also previously been investigated in integer and Abelian fractional quantum Hall states in Ref.~\cite{Oblak2022}, in the context of an infinite cylindrical geometry with an entanglement bipartition perpendicular to the cylindrical axis. There, it was found that the symmetry-resolved entanglement obeyed an area law with subleading corrections in the charge deviations, while the full counting statistics had a Gaussian form. This result can be understood in terms of the Li-Haldane bulk-boundary correspondence~\cite{Li2008,Qi2012,Peschel2011,Chandran2011,Swingle2012} and its irrelevant corrections~\cite{Dubail2012}. In Abelian fractional quantum Hall states such as the Laughlin state considered in Ref.~\cite{Oblak2022}, however, the different topological sectors can be distinguished by the resolution of the $\mathrm{U}(1)$ charge symmetry alone. This raised the question of how the results would generalize in the case of non-Abelian fractional quantum Hall states where this straightforward relation between the topological sectors and the $\mathrm{U}(1)$ charge symmetry is not satisfied. Additionally, in the study of realistic non-Abelian Moore-Read fractional quantum Hall droplets, distinct velocities of neutral and charged modes along the physical edge are observed \cite{Hu2009}, a phenomenon not found in the Laughlin case that presents another complication.

To address these questions, we consider in this work the bosonic Moore-Read (MR) state, a non-Abelian fractional quantum Hall state. We are able to fully resolve the symmetry by accounting for the role of fermion parity symmetry as well as the $\mathrm{U}(1)$ charge symmetry. When properly accounting for fermionic parity, we find similar results for the symmetry-resolved entanglement and full counting statistics within the Abelian and non-Abelian topological sectors, which we numerically confirm using Matrix Product States (MPS) simulations. We additionally explore the corrections to the Li-Haldane conjecture directly by computing a synthetic entanglement Hamiltonian with the first few corrections, which we then fit to the MPS numerical entanglement spectrum. This allows us to gain a more granular understanding of symmetry-resolved entanglement in the bosonic Moore-Read state. Both the MPS and synthetic entanglement Hamiltonian approaches also enable controlled calculation of the distinct charged and neutral (fermionic) velocities.

The paper is organized as follows. In Section \ref{sec:symresrdm} we define notation and various ways of quantifying symmetry-resolved entanglement. 
Then, in Section \ref{sec:sremooreread}, we describe the setup of the bosonic Moore-Read state on a cylinder, starting with the description of the conformal field theory (CFT) of the bosonic Moore-Read state and its connection, via the Li-Haldane correspondence, to the entanglement spectrum. This is then applied in Section \ref{sec:applylh} to describe the picture of how symmetry-resolved entanglement manifests in the bosonic Moore-Read state, presenting some analytical results, including the form of the symmetry-resolved entanglement and full counting statistics derived from the Li-Haldane form of the entanglement Hamiltonian. Section \ref{sec:applylh} also includes our description of the corrections to the Li-Haldane description that will go into the construction of the synthetic entanglement spectrum. 
This is followed by Section \ref{sec:numericalresults}, in which we examine the symmetry-resolved entanglement of the bosonic Moore-Read state numerically, via MPS methods, and in which we demonstrate the additional information we can gain from the synthetic entanglement spectrum approach. Finally, Section \ref{sec:conclusion} summarizes our conclusions, and the appendices contain some additional results, details on the fitting approach used for the synthetic entanglement spectrum, and information about our conventions.

\section{Symmetry-resolved reduced density matrix}
\label{sec:symresrdm}

In this section, we outline notation and describe symmetry-resolved entanglement and related concepts.
A many-body quantum system can be partitioned into spatial regions $A$ and $B$, such that the Hilbert space $\mathcal{H}$ factorizes as $\mathcal{H} \cong \mathcal{H}_A \otimes \mathcal{H}_B$. $\mathcal{H}_A$ and $\mathcal{H}_B$ denote the Hilbert spaces associated with regions $A$ and $B$, respectively. Given a pure state $\ket{\Psi}$, the entanglement between regions $A$ and $B$ is encoded in the reduced density matrix (RDM) $\rho_A \equiv \Tr_B(\rho)$, the trace of the total density matrix $\rho = \ket{\Psi} \bra{\Psi}$ over the degrees of freedom corresponding to $\mathcal{H}_B$, which will yield a density matrix of states in $\mathcal{H}_A$. 

We consider the case where the system satisfies a global symmetry with some locally conserved charge $Q$, i.e., $[\rho,Q] = 0$. This charge can be decomposed according to the spatial partition as $Q = Q_A \otimes \mathbb{I}_B + \mathbb{I}_A \otimes Q_B$, where $Q_A$ and $Q_B$ are the total charges of regions $A$ and $B$, respectively. 
The reduced density matrix $\rho_A$ commutes with $Q_A$ and is thus block-diagonal with respect to eigenspaces of $Q_A$:
\begin{equation}
    \label{eq:rhoadecomposition}
	\rho_A = \bigoplus_q p_{q} \rho_A(q) = \begin{pmatrix}
		\ddots   & & \\
		& p_{q} \rho_A(q) & \\
		& & \ddots 
	\end{pmatrix},
\end{equation}
where the direct sum runs over all eigenvalues $q$ of $Q_A$. The \emph{symmetry-resolved reduced density matrix} $\rho_A(q)$, normalized such that $\Tr_{A} [\rho_A(q)] = 1$, corresponds to the physical reduced density matrix conditioned on a measurement outcome $Q_A = q$. The corresponding weight $p_q$ is the probability of observing this charge sector, and the collection $\{p_q\}$ defines the FCS~\cite{Levitov1993,Levitov1996}. We can write down measures of entanglement for the symmetry-resolved reduced density matrix $\rho_A(q)$, such as symmetry-resolved von Neumann entanglement entropy $S_1(q)$ and symmetry-resolved R\'{e}nyi entanglement entropies $S_n(q)$ (for $n > 1$):
\begin{align}
	S_1(q) &\equiv -\Tr[\rho_A(q)\log\rho_A(q)],&\textrm{ and }  \\
	S_n(q) &\equiv \frac{1}{1-n}\log\Tr\left[\rho_A(q)^n\right], &n>1,
\end{align}
respectively. 

From Eq.~\eqref{eq:rhoadecomposition}, we can also calculate the von Neumann entanglement entropy of the overall reduced density matrix, which yields~\cite{Nielsen2010,Lukin2019}
\begin{align}
    \label{eq:total_entropy_sum_rule}
	S_1 &= -\Tr[\rho_A\log\rho_A] \nonumber \\
    &= - \sum_q p_{q}\log p_{q}+ \sum_q p_{q}S_1(q)\\
	& \equiv S^{\textrm{number}} + S^{\textrm{configuration}}, \nonumber
\end{align}
where the first term, known as the number entropy $S^{\textrm{number}}$, is the Shannon entropy of charge fluctuations within the region $A$, while the second term, known as the configuration entropy $S^{\textrm{configuration}}$, contains the actual contributions to the von Neumann entanglement entropy from the symmetry-resolved entanglement entropies $S_1(q)$ within each sector, weighted by the full counting statistics $p_{q}$. 

To facilitate the computation of symmetry-resolved entanglement entropies, it is useful to introduce a set of quantities known as charged moments~\cite{Goldstein2018} or charged R\'{e}nyi entropies~\cite{Belin2013,Pastras2014,Belin2015,Matsuura2016,Caputa2016,Dowker2016,Shiba2017,Dowker2017}. 
If $Q$ is the conserved charge of a $\mathrm{U}(1)$ symmetry, then the charged moments $\widehat{Z}_n(\alpha)$ are defined as
\begin{equation}
    \label{eq:hatzn}
	\widehat{Z}_n(\alpha) \equiv \Tr\left(e^{i\alpha Q_A}\rho_A^n\right).
\end{equation}
If $Q_A$ has eigenvalues $q \in \mathbb{Z} + \delta$, for some constant real shift $\delta$, then we have the periodicity relation
\begin{equation}
	\widehat{Z}_n(\alpha + 2\pi) =  e^{2\pi i \delta}	\widehat{Z}_n(\alpha) \,.
\end{equation}
Symmetry-resolved entanglement entropies and the FCS can be readily obtained from the Fourier modes of the charged moments 
\begin{equation}
    \label{eq:znq}
	Z_n(q) \equiv \int_{-\pi}^\pi \frac{d\alpha}{2\pi} e^{-i\alpha q} \widehat{Z}_n(\alpha) =  \Tr(\Pi_q \rho_A^n)
\end{equation}
where $\Pi_q$ is the orthogonal projector eigenspace of $Q_A$ with eigenvalue $q$, namely
\begin{equation}
	\Pi_q = \int_{-\pi}^\pi \frac{d\alpha}{2\pi} e^{i\alpha(Q_A-q)} \,. 
\end{equation}
In particular the full counting statistics is recovered as 
\begin{equation}
    \label{eq:pq}
	Z_1(q) =  \Tr(\Pi_q \rho_A) = p_q,
\end{equation}
We can write the symmetry-resolved entanglement entropies in terms of the $Z_n(q)$ as well:
\begin{align}
	S_n(q) &= \frac{1}{1-n} \log \frac{Z_n(q)}{[Z_1(q)]^n}, &n>1, & &\textrm{ and}\\
	S_1(q) &=  -\left. \frac{d}{dn} \frac{Z_n(q)}{[Z_1(q)]^n} \right|_{n=1},
\end{align}
where the relation for the von Neumann symmetry-resolved entanglement entropy can be obtained from that for the R\'{e}nyi symmetry-resolved entanglement entropy.
Below, in Section \ref{sec:applylh}, we will discuss how to go about computing these quantities. But first we discuss the setup in which we wish to do so: the bosonic Moore-Read state.

\section{The bosonic Moore-Read state on a cylinder}
\label{sec:sremooreread}

\subsection{Conformal field theory of the bosonic Moore-Read state}
\label{sec:bosonicmrcft}
 
Certain model wavefunctions for fractional quantum Hall states can be constructed as conformal blocks of a chiral (1+1)-dimensional rational conformal field theory (RCFT)~\cite{Moore1991,Cristofano1991coulomb}. These wavefunctions are believed to capture the topological properties of an effective (2+1)-dimensional topological quantum field theory (TQFT) that describes the universal, long-distance behavior of the system. For the bosonic Moore–Read state, the relevant conformal blocks come from the chiral $\mathrm{SU}(2)_2$ Wess–Zumino–Witten (WZW) model~\cite{Moore1991}.

Thus, on our way to understanding the entanglement properties of the bosonic Moore-Read state, we must first review some of the features of this CFT. The $\mathrm{SU}(2)_2$ WZW theory can be constructed using free fields, namely a Majorana fermion $\psi$ and a compact boson $\varphi$~\cite{Milovanovic1996}. Further, under the assumption that these modes have the same velocity, the effective field theory of the edge of the bosonic Moore-Read state will also be given by the chiral $\mathrm{SU}(2)_2$ WZW theory~\cite{Wen1991}. A few elementary facts about these free fields, useful to fix notations, can be found in Appendix \ref{app:freefields}.

\medskip
In terms of the bosonic field $\varphi$ and the Majorana fermion $\psi$, three currents are constructed:
\begin{equation}
    \label{eq:su22bflink}
	J^{\pm} =  \psi \otimes e^{\pm i\varphi}, \textrm{ and } J^0 =  i \partial \varphi \,.
\end{equation}
Their modes $J^a_n$ satisfy the affine $\mathrm{SU}(2)_2$ algebra, which has three integrable representations with $l=0$, $1$, and $2$. In the sector associated to each $l$, the ground states, which are the highest-weight states, form an isospin $l/2$ multiplet with conformal dimension
\begin{equation}
    \label{eq:hlconformaldim}
    h_l = \frac{l(l+2)}{16}.	
\end{equation} 
The sectors associated to $l=0$, $1$, and $2$ correspond to the three topological sectors $\mathcal{H}_{\mathds{1}}$, $\mathcal{H}_{\sigma}$, and $\mathcal{H}_{\psi}$ of the $\nu = 1$ bosonic Moore-Read theory, where the subscripts denote the anyonic labels:
$\mathds{1}$ (the vacuum), $\sigma$, and $\psi$, respectively.
The highest-weight states in each sector can be expressed in terms of the bosonic field $\varphi$ and the Majorana fermion $\psi$ as follows. 
\begin{itemize}
    \item In the vacuum sector $\mathcal{H}_{\mathds{1}}$ the singlet highest-weight state is the vacuum $\ket{0}$, which corresponds to the identity operator via the state-operator correspondence.
    \item In the $\sigma$ sector $\mathcal{H}_\sigma$ the doublet corresponds to the operators $\left( \sigma \otimes e^{ i \varphi/2}, \sigma \otimes e^{ -i \varphi/2} \right)$. 
    \item In the $\psi$ sector $\mathcal{H}_\psi$ we have the triplet $\left( e^{i \varphi},  \psi ,  e^{-i \varphi} \right)$. 
\end{itemize}

The full topological sectors $\mathcal{H}_a$ are obtained by repeated action of the currents. A more explicit description in terms of the bosonic and fermion Fock spaces can be found in Appendix \ref{app:freefields}.

\subsection{A flux-threaded cylinder}

\begin{figure}
    \centering
    \includegraphics[width=\linewidth]{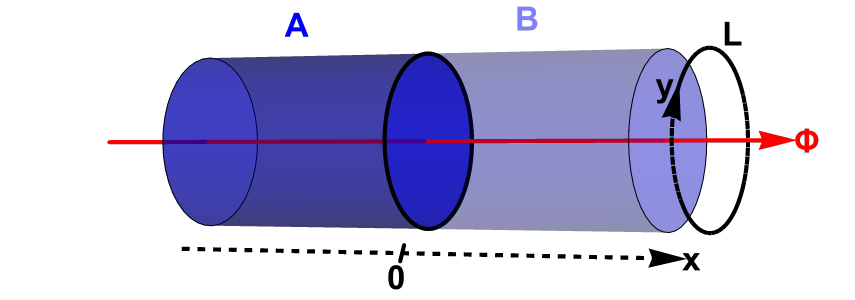}
    \caption{The geometry that we consider is an infinite cylinder of circumference $L$, with  coordinate $x$ in the direction parallel to the axis and coordinate $y$ periodic around the circumference. The entanglement bipartition is between regions $A$ and $B$, where region $A$ covers $x < 0$, while the region $B$ covers $x \geq 0$. Finally, we consider an Aharonov-Bohm flux $\Phi$ threading the cylinder, as shown.}
    \label{fig:cylindergraphic}
\end{figure}

We consider the bosonic MR state on a flat cylinder $\mathbb{R} \times S^1$, where each point is labeled by coordinates $(x, y)$, with the $y$-direction compactified as $y \sim y + L$. We work in units where the magnetic length is set to one, so that the cylinder's perimeter $L$ is expressed in units of the magnetic length (see Fig.~\ref{fig:cylindergraphic}). Using the Landau gauge $\mathbf{A} = x \, d y$, a convenient basis of the lowest Landau level consists of the localized wavefunctions 
\begin{equation}
\label{eq:orbitals}
    \phi_m(x,y) \propto
    e^{ik_my} e^{-(x-k_m)^2/2}, 
    \quad k_m  = \frac{2\pi}{L} (m + \Phi),\,
\end{equation}
where $m \in \mathbb{Z}$, and the parameter $\Phi$ denotes an Aharonov-Bohm flux threading the cylinder. It acts as a tunable external parameter that shifts the allowed momenta $k_m$.

\medskip
We divide the cylinder into two subregions, $A$ and $B$. The region $A$ corresponds to the left ``half-cylinder," defined by $x < 0$, while $B$ is its complement (see Fig.~\ref{fig:cylindergraphic}). Symmetry-resolved entanglement entropies are defined relative to the charge fluctuation in subregion $A$.
\begin{equation}
    \label{eq:qa}
	Q_A = : N_A : - \langle :N_A:\rangle.
\end{equation}
where $N_A$ formally counts the total particle number in $A$, and normal ordering is employed to subtract a divergent contribution that arises because the unbounded region $A$ contains, on average, an infinite number of particles. From the argument presented in~\cite{Oblak2022}, it is expected that, for the $\nu = 1$ MR state in the topological sector $a$,  
\begin{equation}
    \label{eq:NAev}
	Q_A \in \mathbb{Z} + \delta_a(\Phi), \qquad \delta_a(\Phi) = \Phi + q_a + \cdots,
\end{equation}
where $\cdots$ denote small corrections, in the sense that they vanish exponentially as the perimeter $L$ increases, and $q_a$ is the fractional charge of the anyon of type $a$. For the bosonic Moore-Read state under consideration, $q_\mathds{1} = 0$, $q_{\psi}=1$, and $q_\sigma = 1/2$. It is worth noting that, on the infinite cylinder, the distinction between the Abelian topological sectors $\mathds{1}$ and $\psi$ is purely conventional, as they are related by a simple translation---or alternatively, by threading a full unit of flux, $\Phi\to\Phi+1$. In what follows, we adopt the labeling convention consistent with Eq.~\eqref{eq:NAev}.

\subsection{Bulk-edge correspondence}

The entanglement properties of the bosonic MR state in the cylindrical geometry of Fig.~\ref{fig:cylindergraphic} are encoded in the reduced density matrix $\rho_A$ associated with the chosen bipartition. An equivalent and often more insightful perspective is provided by the \emph{entanglement Hamiltonian} 
$H_A$, defined via
\begin{equation}
    \label{eq:HAdef}
    \rho_A \equiv \frac{e^{-H_A}}{Z}, 
\end{equation}
where the denominator $Z = \Tr \, e^{-H_A}$ simply ensures proper normalization. The spectrum of $H_A$, known as the \emph{entanglement spectrum}, plays a crucial role in understanding topological phases. A profound insight by Li and Haldane~\cite{Li2008} revealed a remarkable correspondence: for chiral topological states, the low-lying entanglement spectrum across a given bipartition mirrors the spectrum of the chiral conformal field theory that describes the edge modes induced by a physical cut along the same partition. 

In the geometry shown in Fig.~\ref{fig:cylindergraphic}, we define the entanglement cut along the circle $x=0$, which separates the system into regions $A$ and $B$.
The corresponding conformal Li-Haldane entanglement Hamiltonian will then be
\begin{equation}
    \label{eq:lihaldane}
	H_A = \frac{2\pi v}{L}\left(L_0 - \frac{c}{24}\right),
\end{equation}
where $L_0$ is the zero mode of the CFT stress-energy tensor, $v$ is a non-universal velocity, and $c$ is the CFT central charge. For the $\nu = 1$ bosonic Moore-Read state, $c = 3/2$ and the stress energy tensor is given below in Eq.~\eqref{eq:symmetrict}.

The first subtlety in this picture arises from a careful treatment of the $\mathrm{U}(1)$ charge. In the CFT framework, charge eigenvalues are associated with the eigenvalues of the zero mode $J_0$. In the canonical CFT Hilbert space, as described in Eqs.~\eqref{eq:H0} to \eqref{eq:H1}, the $\mathrm{U}(1)$ charge takes integer values in the Abelian sectors, and half-integer values in the non-Abelian one. However, the physical charge $Q_A$ follows the flux-dependent quantization condition given in Eq.~\eqref{eq:NAev}. For $\Phi=0$, the physical and CFT charge quantizations align perfectly. For other values of the flux, this alignment no longer holds. To reconcile the CFT description with the physical charge spectrum, the CFT Hilbert space must be adjusted by shifting the $\mathrm{U}(1)$ charge accordingly. This shift corresponds to a \emph{spectral flow}, which smoothly interpolates between sectors with different charge quantizations.

An additional complication arises when the bosonic and fermionic modes propagate at different velocities, denoted $v_b$ and $v_f$: 
\begin{equation}
    \label{eq:asymmetriclihaldanesubbed}
    H_A = \frac{2\pi v_b}{L}\left( L_0^{(b)} - \frac{1}{24}\right) + \frac{2\pi v_f}{L}\left( L_0^{(f)} - \frac{1}{48}\right),
\end{equation}
where $L_0^{(b)}-1/24$ and $L_0^{(f)}-1/48$ correspond to the charged/bosonic and neutral/fermionic components of the zero mode of the total stress tensor $T = T^{(b)} + T^{(f)}$, which respectively read:
\begin{align}
    \label{eq:symmetrict}
    T^{(b)} =  \frac{1}{2}:J^2: , \quad T^{(f)} = - \frac{1}{2}:\psi \partial \psi: .
\end{align} 
In the case where $v_b \neq v_f$, the system no longer exhibits $\mathrm{SU}(2)$ symmetry and, importantly, also loses conformal invariance. This stands in contrast to models such as chiral $\mathrm{SU}(2)$ spin liquids, where the $\mathrm{SU}(2)$ symmetry forbids such symmetry breaking~\cite{ArildsenLudwig2022}. One might worry that allowing $v_b \neq v_f$ in the modular Hamiltonian could alter the value of the topological entanglement entropy. As we demonstrate in Appendix \ref{app:modular_TEE}, this is not the case; consequently, such velocity asymmetry should be regarded as a generic feature, consistent with the physical edge \cite{Hu2009}. 

Finally, the most significant complication comes from the fact that one has to include irrelevant perturbations to the above entanglement Hamiltonian, as realized by Dubail, Read, and Rezayi~\cite{Dubail2012}, leading to 
\begin{equation}
	\label{eq:correctedlihaldane}
	H_A =  H_A^{(\textrm{l.o.})} + \sum_{i} g_i \int_0^L \phi_i(y) dy,
\end{equation}
where $H_A^{(\textrm{l.o.})}$ represents the leading order term [given in Eq.~\eqref{eq:asymmetriclihaldanesubbed}], and $\phi_i$ are local irrelevant operators in the CFT, with coupling constants $g_i$. 

The specifics of these corrections, and their impact on the entanglement spectrum, will be discussed in Section \ref{sec:syntheticES}. Their coefficients $g_i$ are non-universal, sensitive to microscopic details and the boundary geometry. Importantly, the associated operators $\phi_i$ are irrelevant in the renormalization group sense; in the geometry considered, they have conformal dimensions $\Delta_i \geq 4$.

In the following section, we analyze the implications of the Li-Haldane form of the entanglement Hamiltonian $H_A$, focusing on its consequences for the full counting statistics, symmetry-resolved entanglement, and entanglement equipartition. We first consider the uncorrected form Eq.~\eqref{eq:asymmetriclihaldanesubbed}, neglecting the subleading terms in Eq.\eqref{eq:correctedlihaldane}, and subsequently incorporate these corrections perturbatively.

\section{Consequences of the Li-Haldane correspondence}
\label{sec:applylh}

\subsection{Leading order}
\label{sec:consequenceslihaldane}

Within the leading approximation of Eq.~\eqref{eq:asymmetriclihaldanesubbed} the entanglement Hamiltonian is quadratic, and the FCS and symmetry-resolved entanglement entropies are straightforward to evaluate (see Appendix \ref{app:modular}). 

At large perimeter $L$, up to exponentially small corrections, all three topological sectors exhibit identical full counting statistics, and there is equipartition of entanglement. The FCS is described by a discrete Gaussian distribution:
\begin{equation}
\label{eq:discrete_gaussian}
    p_q \propto e^{- \frac{q^2}{2\sigma^2}}, \qquad \sigma^2 = \frac{L}{2\pi v_b} \,,
\end{equation}
up to an overall normalization chosen to ensure that the total probability sums to one.
The only distinction between sectors lies in the allowed values of the charge: in the Abelian sectors, $q \in \mathbb{Z}+ \Phi$, whereas in the non-Abelian sector, $q \in \mathbb{Z}+ \Phi + 1/2$. The variance scales with the boundary length (area law), while all higher cumulants are exponentially suppressed in the large-$L$ limit. According to the scaling predictions in Ref.~\cite{Berthiere2025geometric}, odd cumulants are expected to vanish in the considered geometry. In contrast, even cumulants should scale with the perimeter, and the leading approximation to the Li-Haldane Hamiltonian [Eq.~\eqref{eq:asymmetriclihaldanesubbed}] fails to capture this behavior.

At finite perimeter, however, a marked difference arises between the Abelian and non-Abelian sectors. The non-Abelian sector continues to exhibit a purely discrete Gaussian distribution, maintaining exact equipartition of entanglement. The Abelian sectors show a more subtle behavior. While their FCS retains an overall Gaussian envelope, finite-size effects introduce a preference for \emph{even} fermion parity over \emph{odd}. In what follows, we focus on the vacuum sector, noting that the $\psi$ sector shares the same structure and can be obtained by a simple shift $\Phi \to \Phi + 1$.
In the vacuum sector charges corresponding to even fermion parity ($q \in 2\mathbb{Z} + \Phi$) appear with enhanced probability, while those corresponding to odd fermion parity ($q \in 2\mathbb{Z} + \Phi + 1$) are suppressed: 
\begin{equation} 
    \label{eq:paritygaussian}
    p_q \propto  \left( 1 \pm \sqrt{\frac{\theta_4(\tau_f)}{\theta_3(\tau_f)}} \right) e^{- \frac{q^2}{2\sigma^2}},
\end{equation}
where $\tau_f = i v_f / L $, the sign $\pm$ corresponds to fermion parity, and $\theta_3$ and $\theta_4$ are  Jacobi theta functions (see Appendix \ref{app:theta}). These parity effects are controlled by the neutral velocity $v_f$, and they are exponentially suppressed at large perimeter $L$:
\begin{align} 
    \frac{\theta_4(\tau_f)}{\theta_3(\tau_f)} \sim 2\, \exp \left( - \frac{\pi L}{4 v_f} \right), 
\end{align} 
thus recovering the discrete Gaussian distribution of Eq.~\eqref{eq:discrete_gaussian}. 
 
One can quantify the imbalance between even and odd fermion parity sectors by computing the probability of finding $q \in 2\mathbb{Z} + \Phi$
(even fermion parity) and that of having $q \in 2\mathbb{Z} + \Phi + 1$
(odd fermion parity). This corresponds to performing a symmetry resolution with respect to $ \mathbb{Z}_2 $ (fermion parity). At the ``symmetric point" $\Phi = 1/2$, where the FCS distributions of the vacuum and $\psi$ sectors are reflections of one another about $q = 0$, these probabilities take a particularly simple form, which is the same for both vacuum and $\psi$ sectors:
\begin{align}
    \label{eq:evenfcs}
    p_{\mathrm{even}} &= \frac{1}{2} \left( 1 + \sqrt{\frac{\theta_4(\tau_f)}{\theta_3(\tau_f)}} \right), \\ p_{\mathrm{odd}} &= \frac{1}{2} \left( 1 - \sqrt{\frac{\theta_4(\tau_f)}{\theta_3(\tau_f)}} \right), 
    \label{eq:oddfcs}
\end{align}
and the parity imbalance boils down to
\begin{align}
    \label{eq:parityimbalance}
    p_{\mathrm{even}} - p_{\mathrm{odd}} = \sqrt{\frac{\theta_4(\tau_f)}{\theta_3(\tau_f)}} > 0.
\end{align} 
This imbalance is maximal in the thin-torus limit  ($L\to 0$), where $p_{\mathrm{even}} = 1$ and $p_{\mathrm{odd}} = 0$, and vanishes exponentially at large perimeter. 

Having discussed the FCS, we now consider symmetry-resolved entanglement entropies. Just as we write down the modular Hamiltonian $H_A$ in terms of the full reduced density matrix $\rho_A$ [Eq.~\eqref{eq:HAdef}], we can also define the symmetry-resolved modular Hamiltonian $H_{A,a}(q)$ in the topological sector $a$ from the corresponding symmetry-resolved reduced density matrix $\rho_{A,a}(q)$:
\begin{equation}
    \label{eq:HAqdef}
    \rho_{A,a}(q) \equiv \frac{e^{-H_{A,a}(q)}}{{Z_a(q)}},
\end{equation}
where $Z_a(q)$ is the associated normalization factor.

As mentioned previously, in the non-Abelian sector, we have strict equipartition, even at finite perimeter $L$, as it is the case both that the symmetry-resolved modular Hamiltonian does not depend on $q$,
\begin{equation}
    H_{A,\sigma}(q)=\frac{2\pi }{L}\left(v_b\sum_{n=1}^\infty J_{-n} J_n +v_f \sum_{m=1}^\infty m \psi_{-m}\psi_m\right),
\end{equation}
and that the auxiliary CFT Hilbert space on which it acts has the same structure for all $q$'s.

In contrast, at finite $L$, the Abelian sectors exhibit equipartition only within each fermion parity sector. While the formal expression of the symmetry-resolved modular Hamiltonian remains independent of $q$ and the same in both $\mathds{1}$ and $\psi$ sectors and is thus also the same for the cases of both even and odd fermion parity,
\begin{equation}
    H_{A,\mathds{1}/\psi}(q)= \frac{2\pi }{L}\left(v_b\sum_{n=1}^\infty J_{-n} J_n +v_f \sum_{m=1/2}^\infty m \psi_{-m}\psi_m\right)\,,
\end{equation}
the auxiliary CFT Hilbert space on which it acts---and consequently the entanglement spectrum---now depends on the fermion parity
\footnote{Concretely, the the subspace $\ker(J_0 - q)$ is given by $ \mathcal{F}^{(+)}_{\textrm{NS}} \otimes \mathcal{F}_q$ when $q \in 2\mathbb{Z} + \Phi$ (even fermion parity), and $ \mathcal{F}^{(-)}_{\textrm{NS}} \otimes \mathcal{F}_q$ when $q \in 2\mathbb{Z} + \Phi + 1$ (odd fermion parity) in the vacuum sector. In the $\psi$ sector, the structure is analogous up to the aforementioned $\Phi \to \Phi + 1$ shift.}. 

Although the entanglement spectra differ between the even and odd fermion parity sectors, they produce identical entanglement entropies in the large $L$ limit, up to corrections that are exponentially suppressed. This ensures that equipartition of entanglement is restored asymptotically.

\subsection{Corrections to the conformal spectrum}
\label{sec:corrections}

An important question concerning the irrelevant corrections to the Li-Haldane Hamiltonian in Eq.~\eqref{eq:correctedlihaldane} is how they modify the symmetry-resolved entanglement and full counting statistics obtained in the previous subsection. 
We can approach this question by looking at the $\mathrm{U}(1)$-charged moments. Applying Eq.~\eqref{eq:hatzn} to the Li-Haldane equation Eq.~\eqref{eq:lihaldane}, we get that the $\mathrm{U}(1)$-charged moment in the sector $a$ is
\begin{equation}
    \label{eq:lihaldanechargedmoment}
	\widehat{Z}_{n,a}(\alpha) = \frac{1}{Z_a^n} \Tr_a \left[e^{i\alpha Q_A} e^{-H_A}\right]
\end{equation}
Our discussion then follows that of Ref.~\cite{Oblak2022}, interpreting the numerator of the charged moments in Eq.~\eqref{eq:lihaldanechargedmoment} as the partition function of a critical 1D system on an open chain of length $L$ at inverse temperature $\beta_n = 2n$, twisted by an imaginary chemical potential $e^{i \alpha Q_A}$. The quantity $Z_a$ in the denominator corresponds to the same partition function without the twist, computed at $\beta_1$. 
The main departure from the setup in Ref.~\cite{Oblak2022} is the presence of two distinct velocities for the neutral and charged sectors, which breaks rotational invariance in the Euclidean spacetime picture. While we do not account for this subtlety explicitly in the following analysis and proceed heuristically, we expect the resulting scaling behavior to remain valid, and as we will see, our numerical observations in Section \ref{sec:mpsresults} are consistent with this heuristic.

To study the large-$L$ behavior, it is convenient to recast the partition functions in terms of a transfer matrix that evolves along the spatial direction, effectively exchanging space and imaginary time. In this representation, the partition function takes the form
\begin{equation}
    \label{eq:B2B1}
    \Tr_a \left[e^{i\alpha Q_A} e^{-H_A}\right] = \bra{\tilde{0}} e^{-L H^{(\alpha)}_n} \ket{\tilde{a}},
\end{equation}
where $e^{-H^{(\alpha)}_n}$ is the transfer matrix in the presence of the twist $\alpha$, and $\ket{\tilde{0}}$, $\ket{\tilde{a}}$ are the (twisted) boundary states associated to the vacuum and the anyon $a$, respectively, in the sense of Ref.~\cite{Cardy2008boundaryconformalfieldtheory}. In this representation, the twist $\alpha$ is implemented through a spectral flow. 

In the large-$L$ limit, this amplitude is dominated by the ground state $\ket{0}$ of $H^{(\alpha)}_n$, leading to
\begin{equation}
    \label{eq:b1b2factors}
    \bra{\tilde{0}} e^{-L H^{(\alpha)}_n} \ket{\tilde{a}} \sim \braket{\tilde{0}|0}\braket{0|\tilde{a}} e^{-L E_{n}(\alpha)},
\end{equation}
where $E_n(\alpha)$ is the corresponding energy eigenvalue. The overlaps $\braket{\tilde{0}|0}$ and $\braket{0|\tilde{a}}$ are universal and unaffected by the irrelevant corrections~\cite{AffleckLudwig1991}, and they encode the topological entanglement entropy:
\begin{equation}
    \braket{\tilde{0}|0}\braket{0|\tilde{a}}\sim e^{-\gamma_a},
\end{equation}
with $\gamma_a$ the topological entanglement entropy in sector $a$. 
This leads to the following large-$L$ asymptotic form for the charged moments:
\begin{equation}
	\widehat{Z}_{n,a}(\alpha) \sim e^{(n-1) \gamma_a} e^{-L[E_{n}(\alpha)-nE_{1}(0)]}.
\end{equation}
At $\alpha = 0$, this expression reproduces the expected scaling of the total R\'{e}nyi entropies:
\begin{equation}
	S_{n,a} = \frac{1}{1-n} \log \widehat{Z}_{n,a}(0) \sim \frac{E_{n}(0)-nE_{1}(0)}{n-1} L - \gamma_a.
\end{equation}

We now return to the analysis of perturbations to the conformal spectrum. The main observation is that the invariance under $Q_A \mapsto -Q_A$ ensures that $E_n(\alpha)$ is an even function of the twist parameter $\alpha$, and furthermore it exhibits a minimum at zero twist 
\footnote{The invariance under $Q_A \mapsto -Q_A$ allows the partition function in Eq.~\eqref{eq:B2B1} to be rewritten as 
$\mathrm{Tr}_a\!\left[ e^{i\alpha Q_A} e^{-H_A}\right] = \mathrm{Tr}_a\!\left[\cos(\alpha Q_A)\, e^{-H_A}\right]$.
The operator $\cos(\alpha Q_A)$ is Hermitian and satisfies $-1 \leq \cos(\alpha Q_A) \leq 1$. 
Expanding into a simultaneous eigenbasis of $Q_A$ and $H_A$, it follows that $\mathrm{Tr}_a\!\left[ e^{i\alpha Q_A} e^{-H_A}\right]$ reaches a strict maximum at $\alpha =0$. In the Wick-rotated formulation this directly implies $E_n(\alpha) \geq E_n(0)$ with equality  only for $\alpha = 0 \pmod{2\pi}$.}.
Consequently, the large-$L$ asymptotic behavior of the Fourier-transformed partition function $Z_n(q)$ is dominated by the region near $\alpha =0$, allowing us to apply Laplace's method. Expanding around $\alpha =0$, we find
\begin{equation}
    \label{eq:hatZnsigmaexpansion}
	\widehat{Z}_{n,a}(\alpha) \sim e^{(n-1)\gamma_a} e^{-L(a_n+b_n \alpha^2 + c_n \alpha^4+\cdots)},
\end{equation}
which is the same form of the charge moment for the integer and Abelian fractional quantum Hall states described in Ref.~\cite{Oblak2022}, meaning that in the $\sigma$ sector, as well as in the even and odd sectors, the conclusions from Ref.~\cite{Oblak2022} regarding the symmetry-resolved entropies will go through in essentially the same way. In particular, for the symmetry-resolved entropies in these sectors $S_{n,a}$, we will have
\begin{equation}
    \label{eq:snqexpansion}
	S_{n,a}(q) \sim S_{n,a} - \frac{1}{2}\log L + A_n - B_n \frac{q^2}{L} + C_n \frac{q^4}{L^3},
\end{equation}
with the coefficients $A_n$, $B_n$, and $C_n$ given in terms of the $a_n$, $b_n$, and $c_n$ of Eq.~\eqref{eq:hatZnsigmaexpansion} by 
\begin{align}
	A_n &\sim \frac{\log b_n - n \log b_{1}}{2(n-1)} - \frac{\log(4\pi)}{2} - O(1/L),\\ 
	B_n &\sim \frac{n/b_{1} - 1/b_n}{4(n-1)} - O(1/L),\text { and } \\
	C_n &\sim \frac{c_n/b_n^4 - nc_{1}/b_{1}^4}{16(n-1)}.
\end{align}

That said, these coefficients $A_n$, $B_n$, and $C_n$ are unable to be derived analytically, and so we must ultimately obtain them from the numerical data. We do this below in Section \ref{sec:mpsresults}, where we exhibit the results for the symmetry-resolved entanglement from MPS calculation.

In addition, we can also see that the corrections to the Li-Haldane formula will have an effect on the full counting statistics. We have from Eqs.~\eqref{eq:znq} and \eqref{eq:pq} that 
\begin{align}
    p_{q,a} &= Z_{1,a}(q) = \int_{-\pi}^\pi \frac{d\alpha}{2\pi}e^{-i\alpha q}\widehat{Z}_{1,a}(\alpha)   \\
    &\sim \int_{-\pi}^\pi \frac{d\alpha}{2\pi} e^{-i\alpha q}e^{-L(a_{1}+b_{1} \alpha^2 + c_{1} \alpha^4+\cdots)}, \nonumber
\end{align}
but thanks to the quartic terms in $\alpha^4$, this will no longer produce a Gaussian in $q$ as in the Li-Haldane result of Eq.~\eqref{eq:discrete_gaussian}. Indeed, it is from the cumulative effect of all such corrections that one may recover the true FCS, and the scaling of its even cumulants with $L$ predicted by Ref.~\cite{Berthiere2025geometric} that the leading order of Li-Haldane is unable to capture.

In the following subsection, we take a distinct point of view and consider instead how to approach calculation of $S_{n,a}(q)$ by approximating the corrections to the symmetry-resolved entanglement spectra at the level of operators. This can allow us to numerically examine corrections to the symmetry-resolved entanglement and FCS.

\subsection{Constructing a synthetic entanglement spectrum}
\label{sec:syntheticES}

To analyze the corrections to the symmetry-resolved entanglement in more detail, we adopt an approach of building a bottom-up approximation of the entanglement Hamiltonian, creating a ``synthetic" entanglement spectrum with precise control of the perturbative terms that correct Li-Haldane.	
Thus we return to Eq.~\eqref{eq:correctedlihaldane}, which we write in the following way:
\begin{equation}
    \label{eq:lincombhent}
    H_A = \sum_{i=0}^\infty g_i H_A^{(i)}.
\end{equation}
where the $H_A^{(i)}$ are integrals of local chiral boundary operators,
\begin{equation}
    \label{eq:hiintegral}
	H_A^{(i)} = \int_0^L \phi_i(y) dy.
\end{equation}
To see how this this consistent with Eq.~\eqref{eq:correctedlihaldane},
we take $g_0 = v_b$, $\phi_0(y) = (JJ)(y)$, $g_1 = v_f$, and $\phi_1(y) = -(\psi \partial \psi)(y)$. 
Note additionally that conformal scaling guarantees that 
\begin{equation}
    \label{eq:hiscaling}
	H_A^{(i)} = \left(\frac{\pi}{L}\right)^{\Delta_i-1} V_i,
\end{equation}
where $V_i$ is the zero mode of $\phi_i$, for each operator $\phi_i$. For $\phi_0$ and $\phi_1$, we will then have scaling dimension $\Delta_0 = \Delta _1 = 2$. These are thus the operators whose integrals $H_A^{(0)}$ and $H_A^{(1)}$ make up the leading order contribution to the entanglement Hamiltonian. We then have the identification
\begin{align*}
    H_A^{(\textrm{l.o.})} &= g_0 H_A^{(0)} + g_1 H_A^{(1)} \\
    &= \frac{2\pi v_b}{L}\left( L_0^{(b)} - \frac{1}{24}\right) + \frac{2\pi v_f}{L}\left( L_0^{(f)} - \frac{1}{48}\right)
\end{align*}
[cf.~Eq.~\eqref{eq:asymmetriclihaldanesubbed}].
The additional $g_i H_A^{(i)}$ for $i \geq 2$ then serve as the terms that correct $H_A^{(\textrm{l.o.})}$ to get the full $H_A$ in Eq.~\eqref{eq:correctedlihaldane}.

The question then becomes which additional $H_A^{(i)}$ ought to be included in this sum. They should not be the integrals of total derivatives, and they should satisfy the symmetries of the entanglement Hamiltonian. In particular, for integer and half-integer values of $\Phi$, our entanglement Hamiltonian will respect the charge symmetry (coming from the symmetry of the bipartition of the cylinder) $Q_A \mapsto - Q_A$, and hence in the operator language $J \mapsto -J$, so the $H_A^{(i)}$ should contain only an even number of $J$ factors. Another requirement is that the $\phi_i$ have even fermionic parity due to locality considerations.
Accounting for all of these considerations, the most relevant $\phi_i$, with conformal dimension $\Delta_i \leq 4$, whose integrals can occur in the entanglement Hamiltonian are those listed in Table \ref{table:integerevenchargeoperators} \footnote{It is interesting to compare the choice of integrals of local operators we include in the bosonic Moore-Read entanglement Hamiltonian, with those one can choose to include when similarly constructing perturbatively the lower levels of the bosonic Moore-Read edge Hamiltonian, as in Ref.~\cite{Fern2018}}.

\begin{table}
	\centering
	\begin{tabular}{c|c}
		$\Delta_i$ & $\phi_i(y)$ \\
		\hline
		2 &  $(JJ)(y)$, $-(\psi\partial\psi)(y)$ \\ 
		\hline
		\multirow{2}{*}{4} & $(\partial J \partial J)(y)$, $-(\partial \psi  \partial^2 \psi)(y)$, \\
		&   $-((JJ)(\psi \partial \psi))(y)$, $((JJ)(JJ))(y)$ \\
		\hline
	\end{tabular}
	\caption{The most relevant integer-dimensional operators $\phi_i(y)$ with an even number of charge factors $J(y)$ that can appear in the finite-size entanglement spectrum of the bosonic Moore-Read state, sorted by conformal dimension $\Delta_i$.}
    \label{table:integerevenchargeoperators}
\end{table}

Eq.~\eqref{eq:lincombhent} is an infinite sum, but we can truncate it to the most relevant terms (which we take to be the integrals $H_A^{(i)}$ of those operators $\phi_i$ in Table \ref{table:integerevenchargeoperators}), and then we can fit this approximation to the numerical entanglement spectrum from the MPS data of Section \ref{sec:mpsresults}, using the $g_i$ as fitting parameters. The results of these analyses are shown in Section \ref{sec:ESfitresults}.

\section{Numerical results}
\label{sec:numericalresults}

\subsection{MPS results}
\label{sec:mpsresults}

Many different model fractional quantum Hall states on the cylinder, including the bosonic Moore-Read state, can be expressed as exact MPSs~\cite{Zaletel2012,Estienne2013,Estienne2013fractional}. This framework allows for spinful wavefunctions~\cite{Crepel2018,Crepel2019matrix} as well as the presence of quasihole~\cite{Zaletel2012,Estienne2013,Wu2015} and quasielectron~\cite{Kjall2018} excitations. We use the MPS method to compute numerical data for the charge-resolved entanglement spectrum of the bosonic Moore-Read state in all three topological sectors, across a range of system sizes.

In the exact MPS representation of FQH states, the tensors are formally infinite-dimensional as they operate within the Hilbert space of the underlying CFT. However, for practical numerical implementations, these tensors must be truncated. The finite bond dimension of the truncated MPS sets a maximum on the entanglement entropy that the MPS can capture for a cut perpendicular to the cylinder axis (see Fig.~\ref{fig:cylindergraphic}). 
A direct consequence of this truncation is that gapped states (among others), whose entanglement entropy follows an area law and grows proportionally with the cylinder circumference $L$, cannot be faithfully represented in the $L\to \infty$ limit. 

To ensure the reliability of our numerical results, we first determine the range of cylinder perimeters over which the truncated MPS, at the largest truncation parameter considered, accurately captures the Moore-Read state. We use the topological entanglement entropy as a benchmark for this purpose. The details of this process are given in Appendix \ref{app:truncation}. Our analysis shows that the truncation remains valid up to a cylinder perimeter of $L = 12$ magnetic lengths. We therefore restrict our study to the range $8 \leq L \leq 12$, excluding $L < 8$ where the system the Moore-Read state is not yet fully two-dimensional.

\subsubsection{Symmetry resolved entanglement entropy}

\begin{figure*}
    \centering
    \subfloat[\label{fig:MooreRead1comboS2qsubtractedbylihaldaneL8}]{%
      \includegraphics[width=0.6666\columnwidth]{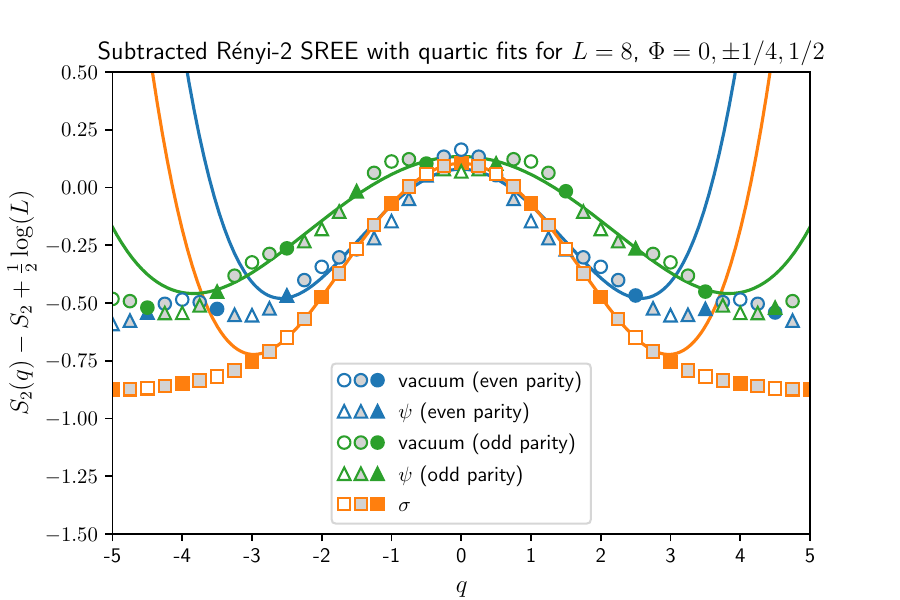}%
    }\hspace*{\fill}%
    \subfloat[\label{fig:MooreRead1comboS2qsubtractedbylihaldaneL10}]{%
      \includegraphics[width=0.6666\columnwidth]{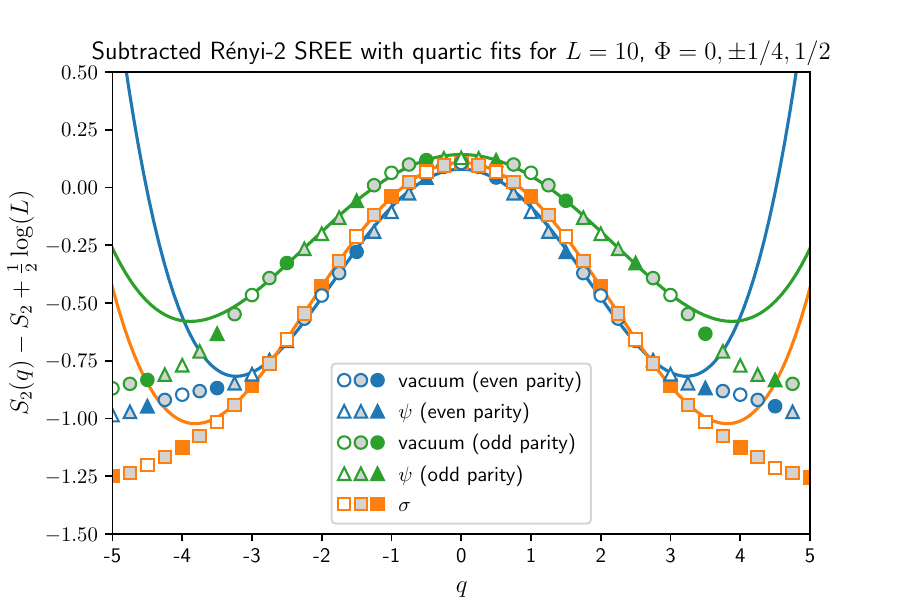}%
    }\hspace*{\fill}%
    \subfloat[\label{fig:MooreRead1comboS2qsubtractedbylihaldaneL12}]{%
      \includegraphics[width=0.6666\columnwidth]{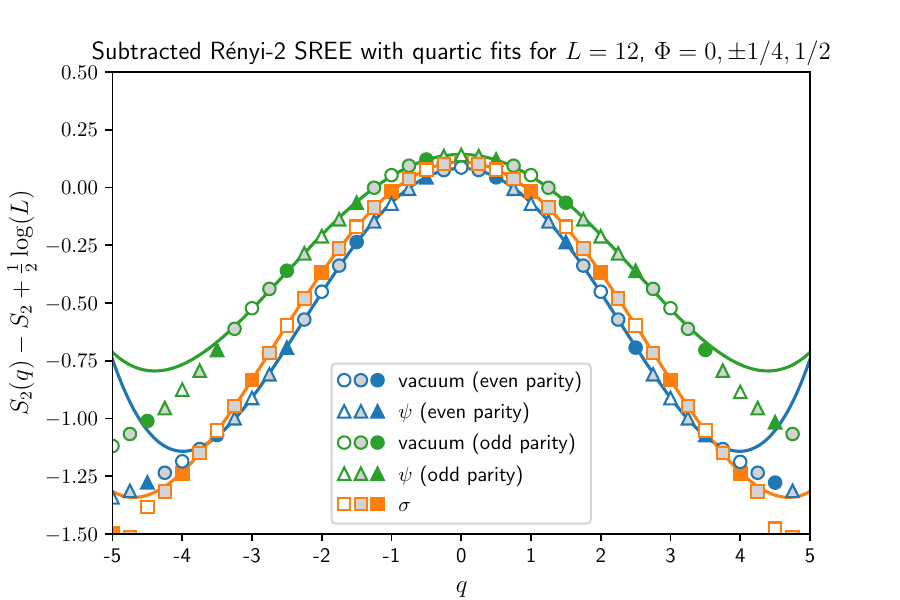}%
    }
    \caption{The subtracted symmetry-resolved second R\'{e}nyi entanglement entropy $S_2(q) -S_1+\frac{1}{2}\log(L) $ for the bosonic Moore-Read state at cylinder perimeter $L=8$ (a), $L=10$ (b) and $L=12$ (c), plotted as a function of the charge $q$, for all three topological sectors. The $\sigma$ sector is plotted in orange. The data for the vacuum and $\psi$ sectors (the circular and triangular markers, respectively) is plotted so as to emphasize the role of even and odd fermionic parity (the blue and green colors, respectively). 
    Data for fluxes $\Phi = 0$, $\pm 1/4$, and $1/2$ is shown by the symbols with white, gray, and filled centers, respectively.
    We also exhibit quartic fits for data of the $\sigma$ sector, and separately for even and odd parity data of the Abelian sectors, for charges $q$ with $q^2 < L$.}
    \label{fig:three_graphs_SREE}
\end{figure*}

\begin{figure*}
    \centering
    \subfloat[\label{fig:MooreRead1comboLiHaldaneA2}]{%
      \includegraphics[width=0.6666\columnwidth]{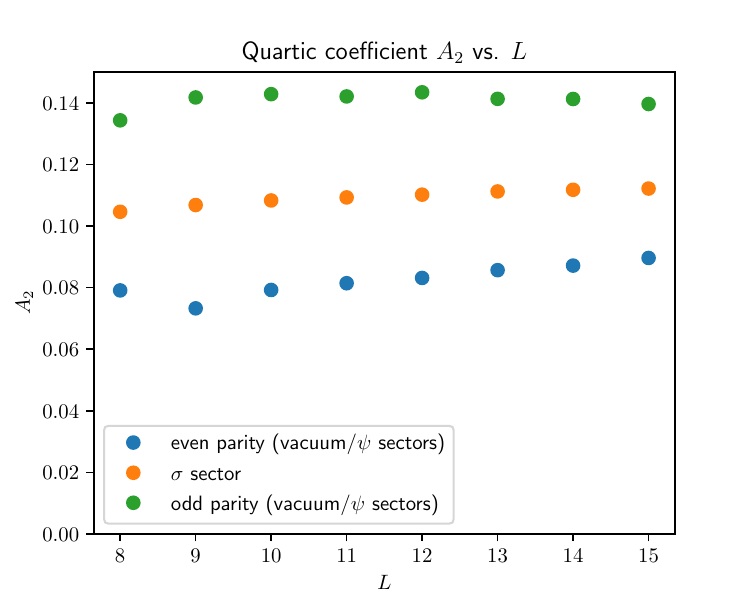}%
    }\hspace*{\fill}%
    \subfloat[\label{fig:MooreRead1comboLiHaldaneB2}]{%
      \includegraphics[width=0.6666\columnwidth]{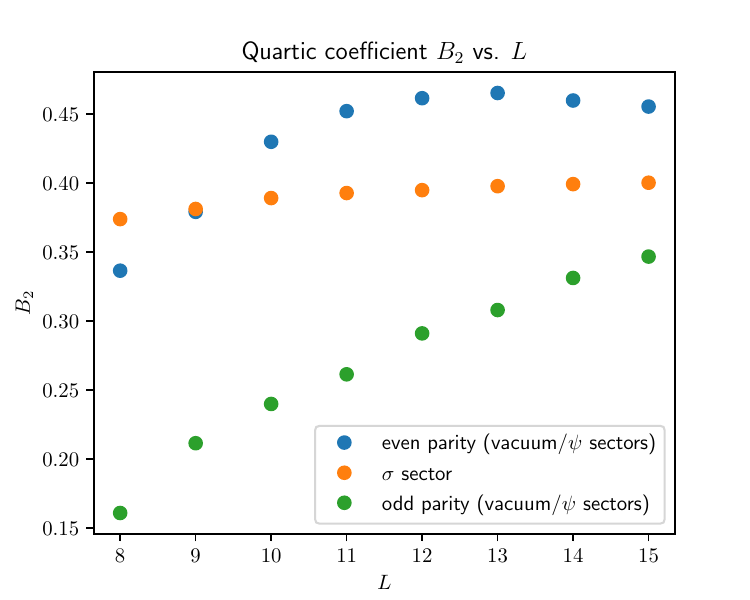}%
    }\hspace*{\fill}%
    \subfloat[\label{fig:MooreRead1comboLiHaldaneC2}]{%
      \includegraphics[width=0.6666\columnwidth]{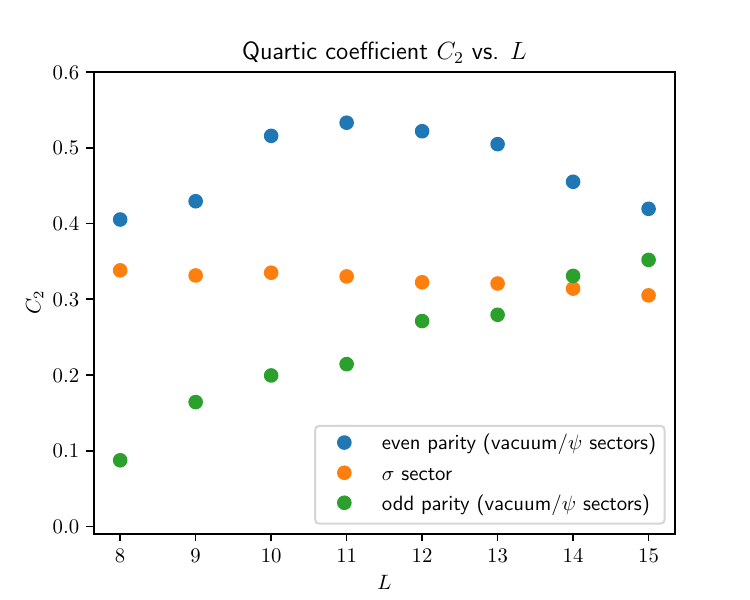}%
    }
    \caption{A plot of the $A_2$ (a), $B_2$ (b), and $C_2$ (c) parameters [see Eq.~\eqref{eq:snqexpansion}] of the quartic fits to the subtracted symmetry-resolved second R\'{e}nyi entanglement entropy $S_2(q) -S_2+\frac{1}{2}\log(L)$, performed to the data of the $\sigma$ sector and the data of even and odd fermionic parity in the Abelian sectors, with $q^2 < L$, as shown in Fig.~\ref{fig:three_graphs_SREE}, versus system size $L$.}
    \label{fig:three_graphs_ABC}
\end{figure*}

We can now analyze the SREE for each system size from the MPS data. At a given system size $L$, we can plot $S_n(q)$. By Eq.~\eqref{eq:snqexpansion}, the subtracted SREE $S_n(q) - S_n + \frac{1}{2} \log L$ can be approximated by a quartic curve in $q$, for the $\sigma$ sector as well as for the odd and even parity ``sectors".
These results are shown for R\'{e}nyi index $n = 2$ in Figs.~\ref{fig:MooreRead1comboS2qsubtractedbylihaldaneL8}, \ref{fig:MooreRead1comboS2qsubtractedbylihaldaneL10}, and \ref{fig:MooreRead1comboS2qsubtractedbylihaldaneL12}, for system sizes $L = 8$, $10$, and $12$, respectively. In these plots, the quartic curves are fit using the data from charges $q$ with $q^2 < L$, consistent with the approximation that went into Eq.~\ref{eq:snqexpansion}. While deviations from the quartic behavior are present at smaller system sizes (and are clearly visible for $L = 8$ in Fig.~\ref{fig:MooreRead1comboS2qsubtractedbylihaldaneL8}), the subtracted SREE values converge well to the quartics already by $L = 12$ in Fig.~\ref{fig:MooreRead1comboS2qsubtractedbylihaldaneL12}. From the quartic fits of the subtracted symmetry-resolved second R\'{e}nyi entanglement entropy at each system size (including those depicted in Fig.~\ref{fig:three_graphs_SREE}), we obtain the coefficients $A_2$, $B_2$, and $C_2$ according to Eq.~\eqref{eq:snqexpansion}, at each system size $L$, in each topological sector. These are plotted in Figs.~\ref{fig:MooreRead1comboLiHaldaneA2}, \ref{fig:MooreRead1comboLiHaldaneB2}, and \ref{fig:MooreRead1comboLiHaldaneC2}. 
The $A_2$, $B_2$, and $C_2$ in each topological sector are expected to converge to a common $A_2$, $B_2$, and $C_2$ at large system size, consistent with the discussion following Eq.~\eqref{eq:b1b2factors}, and indeed, a convergent trend is already apparent for $A_2$ and $B_2$, though less clear for $C_2$ in the TEE-validated system sizes to which we have access. For the Laughlin state SREE in Ref.~\cite{Oblak2022}, it was also harder to resolve the expected trend in $C_2$. This is perhaps to be expected as it is the coefficient of the highest order term in the approximation Eq.~\ref{eq:snqexpansion} and therefore most sensitive to the truncation effects that start to matter more at higher $L$.

\subsubsection{Full counting statistics}

Finally, we look at the FCS. The leading-order Li-Haldane form of the reduced density matrix predicts that the FCS is Gaussian, described by Eqs.~\eqref{eq:discrete_gaussian} for the $\sigma$ sector and \eqref{eq:paritygaussian} for the Abelian sectors at even and odd fermionic parity. We fit our numerical data with Gaussians, which require two parameters along with the circumference $L$: the bosonic velocity $v_b$, which determines the variance $\sigma^2$ of all the Gaussians, 
\begin{equation}
    \label{eq:sigma2L}
    \sigma^2 = \frac{L}{2\pi v_b},
\end{equation}
as seen in Eq.~\eqref{eq:discrete_gaussian}; 
and the fermionic velocity $v_f$, which determines the parity imbalance, and hence the prefactors of the Gaussians for the even and odd fermionic parities in the Abelian sectors.
For all $\Phi$, these prefactors are proportional to the expressions on the right hand sides of Eqs.~\eqref{eq:evenfcs} and \eqref{eq:oddfcs}, which depend on $\tau_f = i v_f/L$ and which, specifically at the $\Phi = 1/2$ symmetric point, describe the $p_{\textrm{even}}$ and $p_{\textrm{odd}}$ of the FCS, as well.
We can perform a single fit to the MPS data simultaneously across all sectors, fluxes $\Phi = 0$, $\Phi = \pm 1/4$, and $1/2$, and the range of TEE-validated circumferences $8\leq L \leq 12$.

\begin{figure*}
    \centering
    \subfloat[\label{fig:MooreRead1FullFCSL8MPS}]{%
      \includegraphics[width=0.6666\columnwidth]{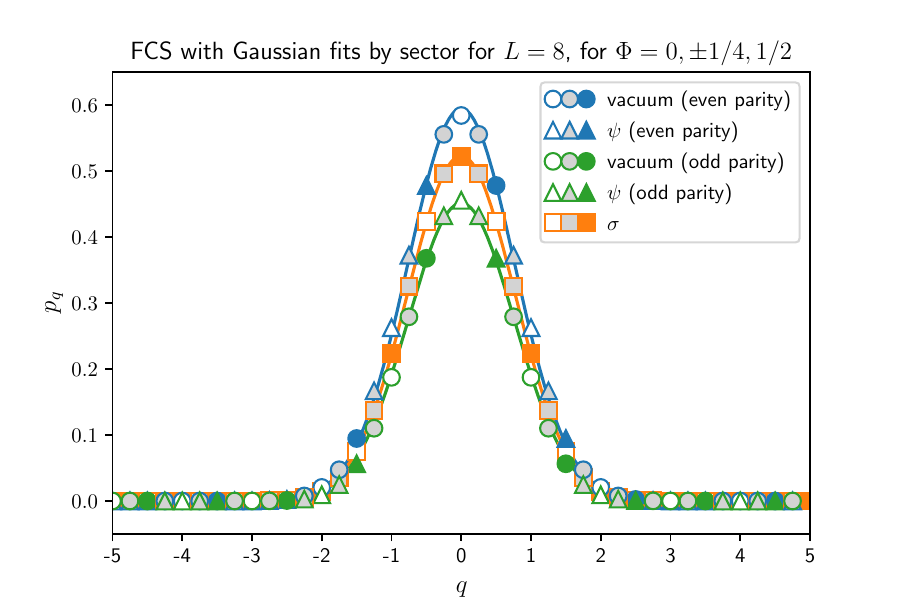}%
    }\hspace*{\fill}%
    \subfloat[\label{fig:MooreRead1FullFCSL10MPS}]{%
      \includegraphics[width=0.6666\columnwidth]{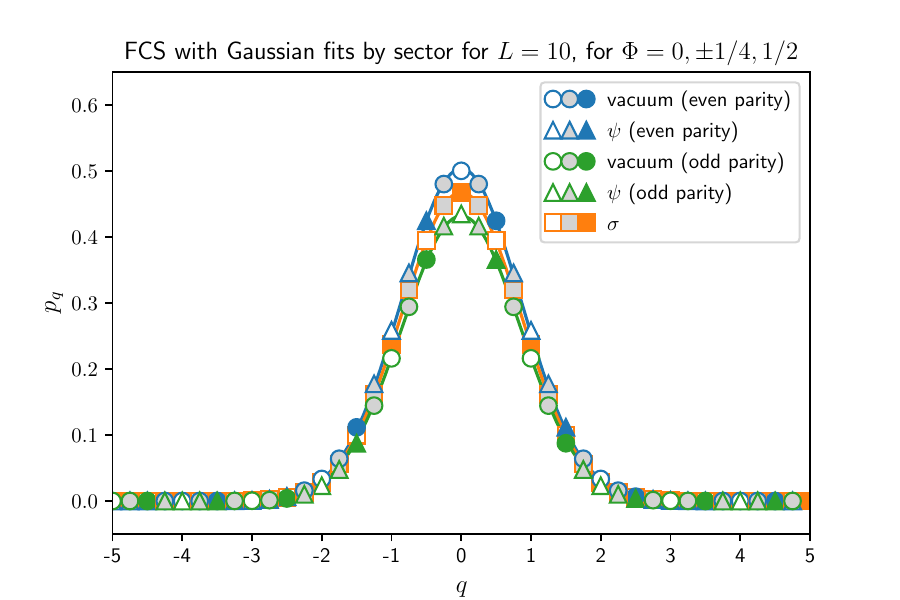}%
    }\hspace*{\fill}%
    \subfloat[\label{fig:MooreRead1FullFCSL12MPS}]{%
      \includegraphics[width=0.6666\columnwidth]{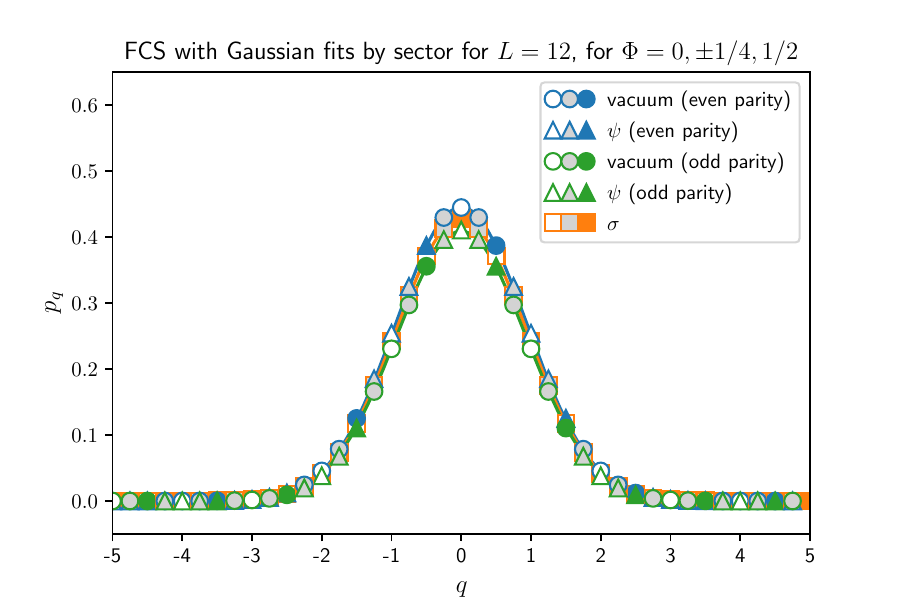}%
    }
    \caption{The FCS for the bosonic Moore-Read state at $\Phi = 0$ (open markers), $\Phi = \pm 1/4$ (gray markers), and $\Phi = 1/2$ (filled markers) plotted as a function $p_{q}$ of the charge $q$, for cylinder perimeter $L = 8$ (a), $10$ (b) and $12$ (c), from the MPS data. The orange square plot markers indicate the FCS from the non-Abelian $\sigma$ topological sector, while the circular and triangular plot markers indicate the vacuum and $\psi$ topological sectors, respectively, with blue and green coloration indicating the even and odd fermionic parities. At each charge associated with the Abelian sectors, it is clear that the even fermionic parity has a higher probability than the odd fermionic parity. The associated curves are the Gaussians from the simultaneous fit to the FCS of all sectors from $L = 8$ through $L = 12$. The corresponding bosonic and fermionic velocities are $v_{b,\textrm{FCS}} \approx 2.19$ and $v_{f,\textrm{FCS}} \approx 1.34$.}
    \label{fig:three_graphs_fcs}
\end{figure*}

We exhibit the FCS at sizes $L = 8$, $L = 10$, and $L = 12$, together with the corresponding Gaussians from the fit, in Figs.~\ref{fig:MooreRead1FullFCSL8MPS}-\ref{fig:MooreRead1FullFCSL12MPS}. This fit finds parameter values of $v_{b,\textrm{FCS}} \approx 2.19$ and $v_{f,\textrm{FCS}} \approx 1.34$ for the bosonic and fermionic velocities, respectively. These can also be compared with the $v_{b,\textrm{FCS}}$ and $v_{f,\textrm{FCS}}$ obtained by fitting the FCS of all sectors and fluxes $\Phi = 0$ and $1/2$ at each circumference $L$ individually. This is done in Appendix \ref{app:vbvfcomparison}. Another possible analysis of the FCS, which provides a way to compute $v_b$, is to analyze its cumulants. This is carried out in Appendix \ref{app:cumulants} and yields similar estimates for the velocities.

\subsubsection{Parity imbalance}

\begin{figure}
    \centering
    \includegraphics[width=\linewidth]{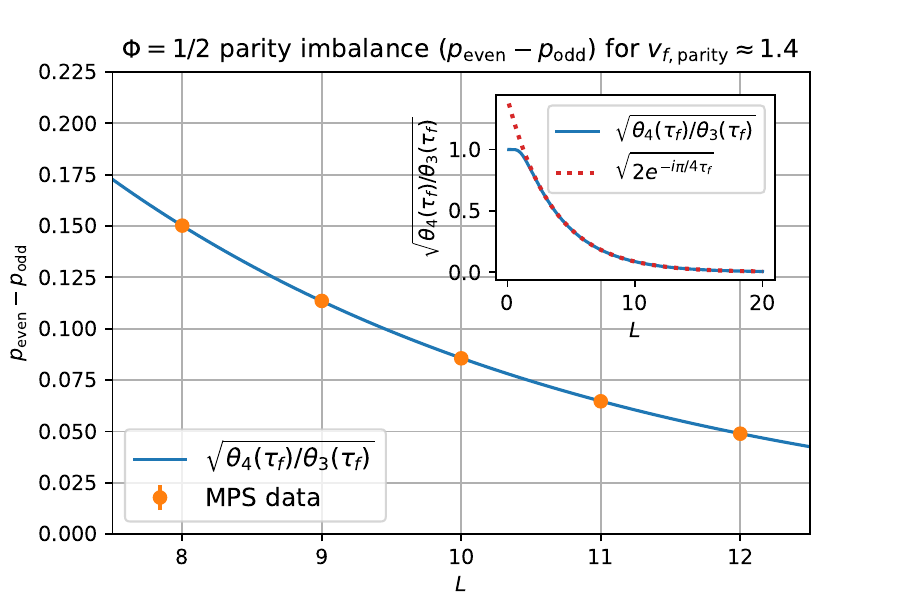}
    \caption{The parity imbalance ($p_{\textrm{even}} - p_{\textrm{odd}}$)
    of the MPS data, averaged over both the vacuum and $\psi$ topological sectors at flux $\Phi = 1/2$ is shown for integer cylinder circumferences $L = 8$ through 12, in the orange points. The blue curve is a fitting of the analytic form $\sqrt{\theta_4(\tau_f)/\theta_3(\tau_f)}$ [see Eq.~\eqref{eq:parityimbalance}], with $\tau_f = i v_f/L$, to this data. The single fitting parameter is the neutral velocity $v_f$, and the result of the fit plotted here has $v_{f,\textrm{parity}} \approx 1.4$. The blue curve $\sqrt{\theta_4(\tau_f)/\theta_3(\tau_f)}$ is also shown plotted over a broader range of $L$ in the inset, along with an asymptotic exponential, from which it is apparent that for the range of $L$ in the main plot the blue curve is well within the exponential regime.}
    \label{fig:MooreRead1paritydiffMPS}
\end{figure}

\begin{figure}
    \centering
    \includegraphics[width=\linewidth]{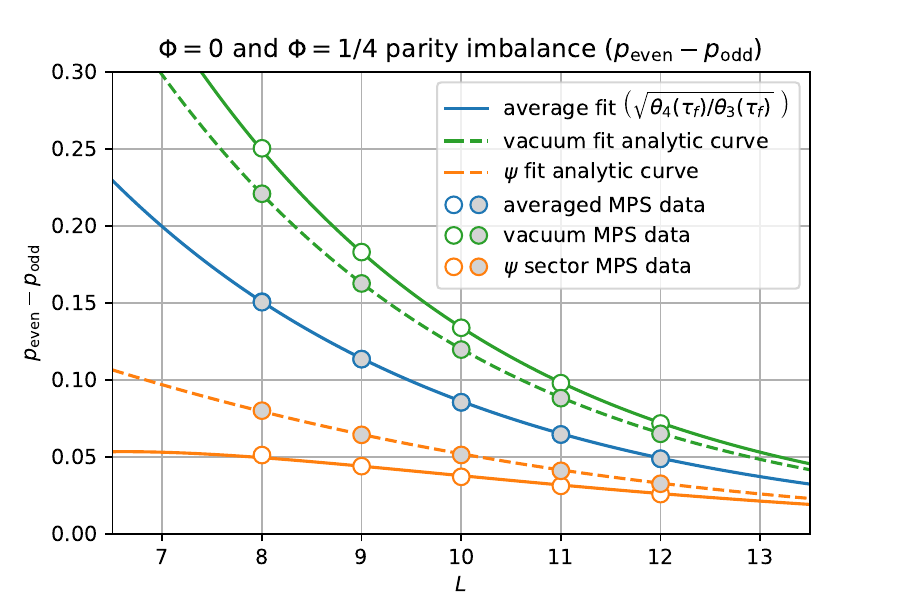}
    \caption{The parity imbalance ($p_{\textrm{even}} - p_{\textrm{odd}}$)
    of the MPS data in both the vacuum and $\psi$ sectors is shown for cylinder circumferences $L = 8$ through 12 at flux $\Phi = 0$ and $\Phi = 1/4$, with the white- and gray-filled markers, respectively. These are fit with analytic curves, parametrized by the bosonic and fermionic velocities $v_b$ and $v_f$, as described in Appendix \ref{app:fcsabelian}. The parameter values obtained are $v_{b,\textrm{parity}} \approx 2.11$ and $v_{f,\textrm{parity}} \approx 1.40$. The average parity imbalances of the vacuum and $\psi$ sectors are also shown for both values of the flux, which overlap almost exactly, and lie along the blue curve, which is parametrized solely in terms of the fermionic velocity $v_f$: $\sqrt{\theta_4(\tau_f)/\theta_3(\tau_f)}$, where $\tau_f = i v_f/L$.}
    \label{fig:MooreRead1paritydiffcomparativeMPS}
\end{figure}

We can also extract Li-Haldane estimates for $v_b$ and $v_f$ from the parity imbalance of the FCS, $p_{\textrm{even}}-p_{\textrm{odd}}$. The analytic form of the parity imbalance is given for both vacuum and $\psi$ sectors by Eq.~\eqref{eq:parityimbalance} at the symmetric point $\Phi = 1/2$, when we do not consider corrections to the Li-Haldane reduced density matrix. The analytic form in this symmetric case can be parametrized by the fermionic velocity $v_f$ alone. In Fig.~\ref{fig:MooreRead1paritydiffMPS}, we plot the parity imbalance of the MPS data at $\Phi = 1/2$ at circumferences $L = 8$ through $L = 12$ along with the analytic curve, which is fit to the data using $v_f$ as the single fitting parameter. This gives a result of $v_{f,\textrm{parity}} \approx 1.4$, consistent with our previous approach. Away from $\Phi = 1/2$, the parity imbalance becomes more intricate and encodes additional information. General expressions valid for arbitrary flux are provided in Appendix~\ref{app:fcsabelian}. In this regime, the imbalance differs between the two Abelian sectors and exhibits an oscillatory dependence on the flux, with an amplitude governed by $v_b$, as described asymptotically at large $L$ in Eq.~\eqref{eq:pimbalance_app}.
The $\Phi = 0$ and $\Phi = 1/4$ MPS parity imbalances in both sectors are shown for $L = 8,\ldots,12$ in Fig.~\ref{fig:MooreRead1paritydiffcomparativeMPS}. Fitting the respective analytic expressions to both the vacuum and $\psi$ sector parity difference data of each flux, we obtain values for the bosonic and fermionic velocity of $v_{b,\textrm{parity}} \approx 2.11 $ and $v_{f,\textrm{parity}} \approx 1.40$ for both $\Phi = 0$ and $\Phi = 1/4$. It is also worth noting that the average of the parity imbalances of the vacuum and $\psi$ topological sectors at each flux recovers once again the form of Eq.~\eqref{eq:parityimbalance}, as depicted by the blue curve of Fig.~\ref{fig:MooreRead1paritydiffcomparativeMPS}.

We can now compare these results with some of those for the same quantities, obtained instead from the synthetic entanglement spectrum of Section \ref{sec:syntheticES}, and we will see that we can also account for terms in the Li-Haldane Hamiltonian beyond leading order in the calculation of $v_b$ and $v_f$.

\subsection{Synthetic entanglement spectrum results}
\label{sec:ESfitresults}

There are a number of ways to optimize the parameters $g_i$ of Eq.~\eqref{eq:lincombhent} in order for $H_A$ to match the MPS entanglement Hamiltonian as closely as possible. These and other technical details regarding this fitting method, such as the appropriate weighting procedure, along with the precise form and computation of the $H_A^{(i)}$, are described in Appendix \ref{app:syntheticES}. 
In this section, we present analysis of some results obtained from directly fitting the levels of the entanglement spectrum with the infinite sum of Eq.~\eqref{eq:lincombhent} truncated to only include terms in $H_A^{(i)}$ that are integrals of operators of dimension 4 or lower. The weight of each entanglement spectrum level in the fit is exponentially suppressed at higher entanglement energies, akin to the weighting of the spectrum of $-\log \rho_A$ by that of $\rho_A$ [see Eq.~\eqref{eq:HAdef}] in the expression for von Neumann entanglement entropy.

\begin{figure*}
    \centering
    \subfloat[\label{fig:MooreRead1vacuumTEE1vsLfit}]{%
      \includegraphics[width=0.6666\columnwidth]{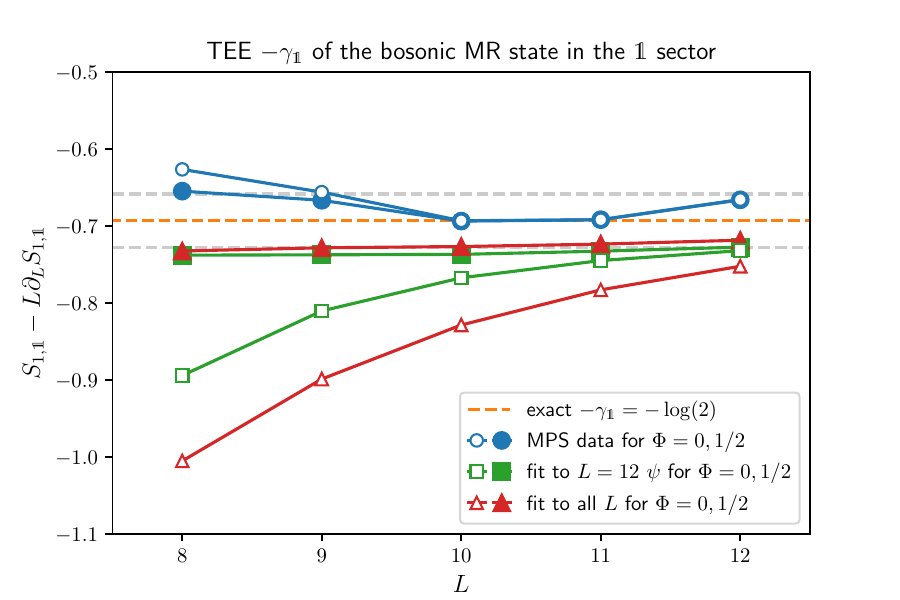}%
    }\hspace*{\fill}%
    \subfloat[\label{fig:MooreRead1sigmaTEE1vsLfit}]{%
      \includegraphics[width=0.6666\columnwidth]{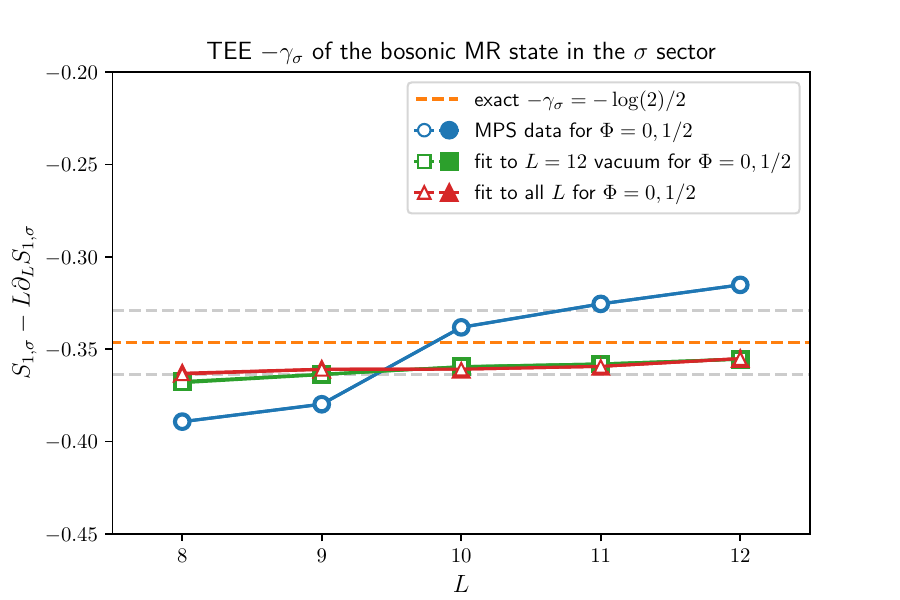}%
    }\hspace*{\fill}%
    \subfloat[\label{fig:MooreRead1psiTEE1vsLfit}]{%
      \includegraphics[width=0.6666\columnwidth]{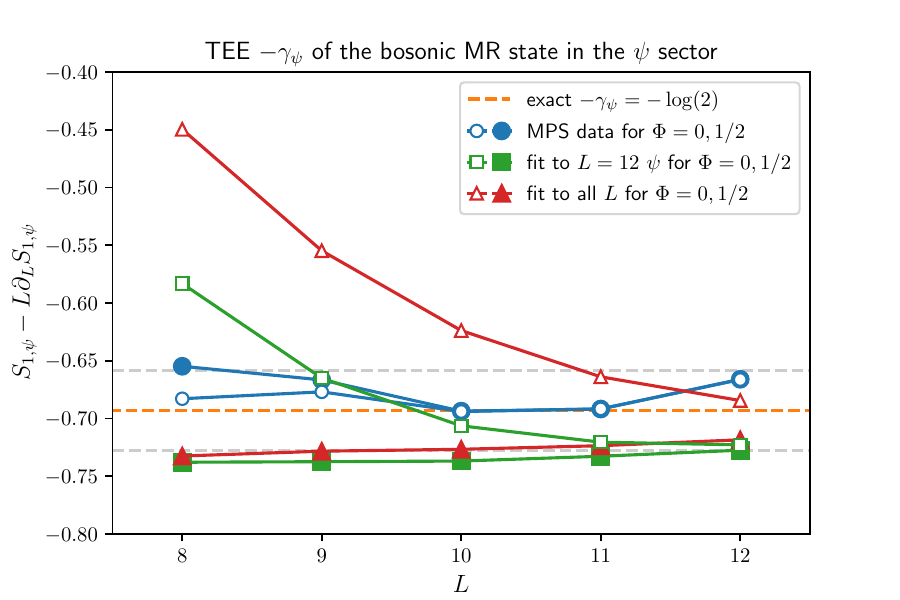}%
    }
    \caption{Von Neumann TEE in each of the vacuum $\mathds{1}$ (a), $\sigma$ (b), and $\psi$ (c) sectors at $\Phi = 0$ and $1/2$ (open and filled markers, respectively), calculated from the von Neumann entanglement entropy $S_{1,a}$ (for the sector $a$) and its derivative with respect to $L$, which is found for a given $L$ using the $S_{1,a}$ at $L\pm 0.01$, plotted as a function of cylinder perimeter $L$ for $L = 8,\ldots,12$. This is done for both the MPS spectrum data (blue circular markers) and the synthetic entanglement spectrum data with two different sets of fitting parameters: those obtained from fitting the $\psi$ MPS data at $L$ = 12 (green squares) and from fitting the $\psi$ and $\sigma$ sectors for all $L$ (red triangles). These may be compared with the plotted exact value (dashed orange line, with dashed gray lines indicating $\pm 5\%$ margins).}
    \label{fig:three_graphs_TEEfit}
\end{figure*}

We perform the fit to the $L = 12$ data in the $\psi$ sector at $\Phi = 1/2$. For most of the system sizes we consider ($L = 8, \ldots, 12$, where the MPS data had a reasonable TEE value), this fitting approach provides a better picture of the TEE of the Abelian sectors, when taking into account both the $\Phi = 0$ and $\Phi = 1/2$ fluxes, than does, for example, fitting to all $L$ and both $\psi$ and $\sigma$ sectors. The relative performance of these fitting approaches in the three sectors for $\Phi = 0$ and $\Phi = 1/2$ can be seen in Fig.~\ref{fig:three_graphs_TEEfit}. We can understand this from the fact that the higher system size MPS data will have less influence from some of the higher-dimensional finite size effects that the synthetic ES cannot capture. Moreover, some effects of fermionic parity in the Abelian sectors will be invisible in the $\sigma$ sector data, so including that data in the fit reduces their salience. The specific parameters $g_i$ of the $L = 12$ $\psi$ fit at $\Phi = 1/2$ used throughout the rest of this section are enumerated in the bold row of Table \ref{table:expsupfitevenchargesdata} of Appendix \ref{app:fittingdata}. The resulting synthetic entanglement spectrum is also graphically compared with that from the MPS data in Appendix \ref{sec:mpsescomparisons}.

We can then reproduce the calculations of the previous subsection, now instead using the ``synthetic" entanglement spectrum generated by the $H_A$ of Eq.~\eqref{eq:lincombhent} with the optimized choice of $g_i$. To simplify our computations, there is an additional truncation (as described in Appendix \ref{app:syntheticES}): we only calculate the synthetic entanglement spectrum to $P_{\max} = 10$ descendant levels above the primary state. However, the restriction of the MPS data to this many descendant levels (from the $P_{\max} = 16$ available in the MPS data) indicates that the decrease in quality is not substantial. This is discussed further in Appendix \ref{app:truncation}. 

A valuable application of the synthetic entanglement spectrum approach is that we are able to get more accurate values for both $v_b$ and $v_f$ that take into account the corrections to Li-Haldane. In particular, these will be given by the parameters $g_0$ and $g_1$, respectively, found by the fit, which can be read off from Table \ref{table:expsupfitevenchargesdata}. 
E.g., for the fit of $L = 12$ $\psi$ sector data that we have considered in this section, we find
\begin{align}
    \label{eq:syntheticvbvf}
    v_b = g_0 \approx 1.82 &\textrm{ and } v_f = g_1 \approx 0.774.
\end{align}
Compared to the estimates obtained above, these values include renormalization effects from all irrelevant terms listed in Table~\ref{table:integerevenchargeoperators}, reducing their sensitivity to finite-size irrelevant corrections, and better capturing the thermodynamic values of the velocities.

\begin{figure}
    \centering
    \includegraphics[width=\linewidth]{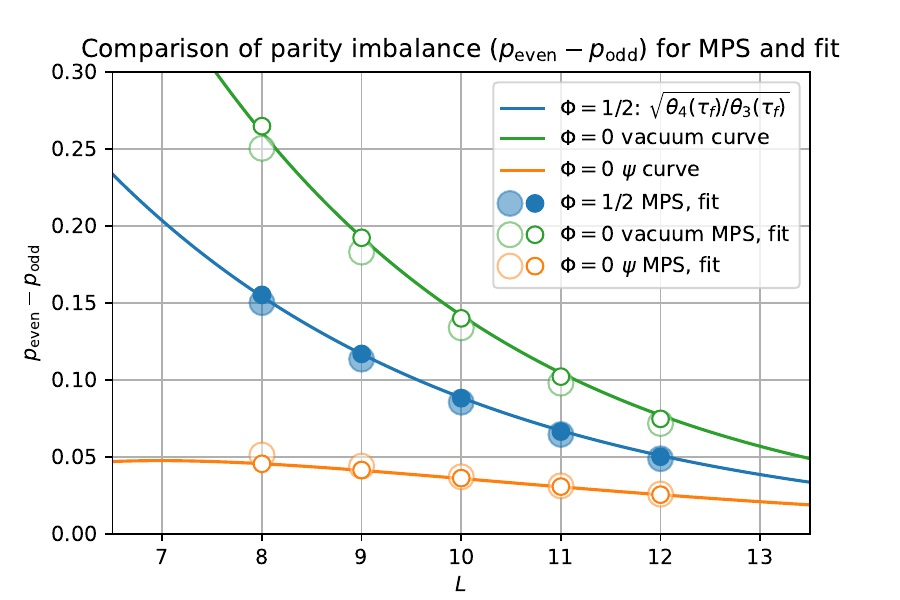}
    \caption{The parity imbalance ($p_{\textrm{even}} - p_{\textrm{odd}}$)
    of the synthetic ES data in both the vacuum and $\psi$ sectors is shown for cylinder circumferences $L = 8$ through 12 at flux $\Phi = 0$ and $\Phi = 1/2$. The corresponding MPS data is shown as well with the larger, faded data points.
    The synthetic ES data parity imbalances in the at $\Phi = 0$ are fit with analytic curves, parametrized by the bosonic and fermionic velocities $v_b$ and $v_f$, as described in Appendix \ref{app:fcsabelian}. The parameter values obtained are $v_{b,\textrm{parity}} \approx 2.17$ and $v_{f,\textrm{parity}} \approx 1.42$. The parity imbalances of the vacuum and $\psi$ sectors at $\Phi = 1/2$ are equal, and are separately fit with the blue curve, which is parametrized solely in terms of the fermionic velocity $v_f$: $\sqrt{\theta_4(\tau_f)/\theta_3(\tau_f)}$, where $\tau_f = i v_f/L$.}
    \label{fig:MooreRead1paritydiffcomparativefit}
\end{figure}

To properly compare this to the Li-Haldane leading order estimates for $v_b$ and $v_f$ that we obtained from the MPS data via the FCS, parity imbalance, and cumulants (see the previous section), we can calculate Li-Haldane leading order estimates for $v_b$ and $v_f$ from these quantities calculated instead from the synthetic entanglement spectrum. For the FCS and second cumulant, this is done in Appendix \ref{sec:mpsescomparisons}. We present the parity imbalance result here, in Fig.~\ref{fig:MooreRead1paritydiffcomparativefit}. The synthetic ES parity imbalance in the vacuum and $\psi$ topological sectors at flux $\Phi = 0$ and system sizes $L = 8, \ldots, 12$ can be fit with analytic curves that are parametrized by $v_b$ and $v_f$, as was done for the MPS data at $\Phi = 0$ and $\Phi = 1/4$ in Fig.~\ref{fig:MooreRead1paritydiffcomparativeMPS}. This gives an estimate for $v_{b,\textrm{parity}} \approx 2.17$ and $v_{f,\textrm{parity}} \approx 1.42$, comparable to the $v_{b,\textrm{parity}}\approx 2.11$ and $v_{f,\textrm{parity}}\approx 1.40$ results from the MPS data 
\footnote{It should also be noted that the slight differences between the synthetic ES and MPS results for $v_{b,\textrm{parity}}$ and $v_{f,\textrm{parity}}$ are not a consequence of the truncation to $P_{\max} = 10$ used in the synthetic ES, as the truncation of the MPS data to $P_{\max} = 10$ produces approximately the same result for $v_{b,\textrm{parity}}$ and $v_{f,\textrm{parity}}$ as the full $P_{\max} = 16$ MPS data, to within the precision quoted here.}.

However, there is some discrepancy between these values and the underlying synthetic ES $v_b$ and $v_f$ of Eq.~\eqref{eq:syntheticvbvf}. This can be mostly accounted for by contributions to the parity imbalance from the integrals of higher-order operators such as $-((JJ)(\psi \partial \psi))(y)$ and $-(\partial \psi  \partial^2 \psi)(y)$. The synthetic ES approach is able to separate out these contributions from the overall effect. As all the $g_i$ are positive, this explains why the $v_b$ and $v_f$ estimates directly from the synthetic ES are lower than those from the parity imbalance. (Although, there will be contributions from operators at dimensions $\Delta_i \geq 6$ that the synthetic ES, truncated at $\Delta_i \leq 4$, cannot isolate either.) A similar story holds for the full FCS fit, as discussed in Appendix \ref{sec:mpsescomparisons}.

\section{Conclusion}
\label{sec:conclusion}

In this work, we considered the symmetry-resolved entanglement of the simplest non-Abelian fractional quantum Hall state, the $\nu = 1$ bosonic Moore-Read state. We verified the presence of approximate entanglement equipartition in the thermodynamic limit, and we observed leading order charge-dependent corrections to the $\mathrm{U}(1)$ symmetry-resolved entanglement entropy [Eq.~\eqref{eq:snqexpansion}]. 
These corrections resemble those previously found for the integer quantum Hall and Laughlin fractional quantum Hall states in Ref.~\cite{Oblak2022}. The main difference is the influence of fermionic parity, which is expected to be generic to all FQH pair states. These results were substantiated with numerical data obtained from an MPS realization of the bosonic Moore-Read wavefunction on the cylinder. 

From this data, we were also able to compute full counting statistics, from which we could extract the distinct bosonic and fermionic velocities of the bosonic Moore-Read state, using the Li-Haldane bulk-boundary correspondence between the entanglement Hamiltonian and the boundary CFT. We further employed the Li-Haldane correspondence, and its corrections, to write down an approximation to the entanglement spectrum, using just the first few integrals of operators more irrelevant than the CFT energy momentum tensor that enter into the entanglement Hamiltonian. This relied upon a set of parameters that we determined by fitting our ``synthetic" entanglement spectrum to the results from the MPS data. The synthetic spectrum provided a good approximation to the MPS results and revealed additional insights into the contributions of CFT operators to the entanglement Hamiltonian. In particular, it enabled a more precise determination of the bosonic and fermionic velocities.

Looking ahead, one additional matter to consider would be resolution with respect to quantities other than $\mathrm{U}(1)$ charge and fermionic parity, such as momentum around the cylinder. Although this work focused on the bosonic Moore-Read state, it represents only one particular model state. Our approach is readily extendable to a broader class of FQH states, including spinful systems that exhibit enhanced global symmetries.

\textit{Acknowledgements}.---B.E.\ acknowledges helpful discussions with J.\ Dubail and Y.\ Ikhlef. N.R.\ and B.E.\ acknowledge B.\ Oblak for collaborations on related topics. M.J.A.\ acknowledges financial support from the PNRR MUR project PE0000023-NQSTI. The Flatiron Institute is a division of the Simons Foundation.

\appendix

\section{Free field CFT relations used in the description of the chiral $\mathrm{SU}(2)_2$ WZW theory}
\label{app:freefields}

{\noindent\textbf{{Free boson.}}\;} The chiral field $\varphi(z)$ represents the holomorphic component of a (compact) free massless boson, normalized such that  
\begin{align*}
    \langle \varphi(z_1) \varphi(z_2) \rangle = - \ln (z_1 - z_2).
\end{align*}
In radial quantization, its mode expansion takes the form
\begin{align*}
    \varphi(z) = \varphi_0 - i J_0 \log (z) + i \sum_{n \neq 0} \frac{1}{n} J_n z^{-n},
\end{align*}
where the modes satisfy the commutation relations
\begin{align*}
    [J_n, J_m] = n \delta_{n+m,0}, \qquad [\varphi_0, J_0] = i.
\end{align*}
The invariance of the theory under shifts $\varphi(z) \to (\varphi(z) +  $constant$)$ leads to the conservation of the holomorphic current 
\begin{align*}
    J(z) = i \partial \varphi(z) = \sum_{n \in \mathbb{Z} } J_n z^{-n-1}.
\end{align*}
As a result, the Hilbert space $\mathcal{H}_{\textrm{boson}} $ is a direct sum of charge sectors $\mathcal{F}_q$ labeled by the eigenvalues $q$ of $J_0$: 
\begin{align*}
    \mathcal{H}_{\textrm{boson}} = \bigoplus_q \mathcal{F}_q\,.
\end{align*}
For a compact boson, these eigenvalues (the $\mathrm{U}(1)$ charge) are quantized depending on the compactification radius. The $\nu = 1$ bosonic Moore-Read theory we consider has compactification radius 1, leading to integer $\mathrm{U}(1)$ charges $q$. Each sector $\mathcal{F}_q$ is naturally organized as a Fock space generated by the creation operators $J_{-n}$ acting on the highest-weight state $\ket{q}$. This chiral CFT has central charge $c_b = 1$, with the stress-energy tensor given by  
\begin{align*}
    T_{\textrm{boson}}(z)  = \frac{1}{2} :J^2(z):.
\end{align*}
The Virasoro generators are quadratic in the modes $J_n$, and in particular the zero mode is 
\begin{align}
    \label{eq:symmetricL0b}
    L^{(b)}_0 = \frac{1}{2} J_0^2 + \sum_{m > 0} J_{-m} J_m.
\end{align}
This operator determines the energy levels, with contributions from both the charge sector (through $J_0$) and the oscillator excitations (through the $J_m$, $m \neq 0$). 

\medskip
{\noindent\textbf{{Majorana fermion.}}\;} The chiral fermion field $\psi(z)$ satisfies the holomorphic OPE:
\begin{align*}
    \psi(z_1) \psi(z_2) \sim \frac{1}{z_1 - z_2}.
\end{align*}
In radial quantization, it admits the mode expansion
\begin{align*}
    \psi(z) = \sum_{n} \psi_n z^{-n - \frac{1}{2}},
\end{align*}
where the modes $\psi_n$ obey the anticommutation relations
\begin{align*}
    \{\psi_n, \psi_m\} = \delta_{n+m, 0}.
\end{align*}
The index $n$ takes different values depending on the boundary conditions imposed on the fermion field, which determine the structure of the Hilbert space. In the {\bf Neveu-Schwarz} (NS) sector the mode indices are half-integers ($n \in \mathbb{Z} + \frac{1}{2}$), while in the  {\bf Ramond} (R) sector they are integers ($n \in \mathbb{Z}$):
\begin{align*}
    \mathcal{H}_{\textrm{fermion}} =  \mathcal{F}_{\textrm{NS}} \oplus  \mathcal{F}_{\textrm{R}} \,. 
\end{align*}
Both sectors are a fermionic Fock space constructed by acting with the creation operators $\psi_{-n}$ ($n > 0$) on a highest-weight state, which is annihilated by all positive modes $\psi_n$ ($n>0$). In the NS sector, this highest-weight state is the vacuum, denoted $\ket{0}$, while in the Ramond (R) sector, it is denoted by $\ket{\sigma}$. The stress-energy tensor of the chiral fermion is
\begin{align*}
    T_{\textrm{fermion}}(z) = -\frac{1}{2} : \psi \partial \psi :(z) ,
\end{align*}
and the central charge of the associated chiral CFT is $c_f = 1/2$.
The corresponding Virasoro generators are given by
\begin{align*}
    L^{(f)}_m = \frac{1}{2} \sum_k \left( k+ \frac{1}{2} \right) :\psi_{m-k} \psi_k: 
\end{align*}
and in particular the Virasoro zero mode  
\begin{align}
    \label{eq:symmetricL0f}
    L^{(f)}_0 = 
    \begin{cases} 
        \displaystyle\sum_{n > 0} n \psi_{-n} \psi_n, & (\textrm{NS :} \,  n \in \mathbb{Z} +1/2) \\[3ex]
        \displaystyle\frac{1}{16}  + \displaystyle\sum_{n > 0} n \psi_{-n} \psi_n , & (\textrm{R :}  \, n \in \mathbb{Z}) 
    \end{cases}
\end{align}
determines the energy levels in each sector.

\medskip
{\noindent\textbf{Fock space construction of the topological sectors.}\;} 
In terms of the bosonic and fermionic Fock spaces, the Abelian topological sectors are constructed out of the Neveu-Schwarz sector and integer $\mathrm{U}(1)$ charges as follows:
\begin{align}
    \label{eq:H0}
    \mathcal{H}_{\mathds{1}} & =  \mathcal{F}^{(+)}_{\textrm{NS}} \otimes \left( \bigoplus_{q \textrm{ even}} \mathcal{F}_{q} \right) \, \oplus \, \mathcal{F}^{(-)}_{\textrm{NS}} \otimes \left( \bigoplus_{q \textrm{ odd}} \mathcal{F}_{q} \right) \\
    \mathcal{H}_\psi & =  \mathcal{F}^{(-)}_{\textrm{NS}} \otimes \left( \bigoplus_{q \textrm{ even}} \mathcal{F}_{q} \right) \, \oplus \, \mathcal{F}^{(+)}_{\textrm{NS}} \otimes \left( \bigoplus_{q \textrm{ odd}} \mathcal{F}_{q} \right)
    \label{eq:H2}
\end{align}
where $ \mathcal{F}^{(+)}_{\textrm{NS}}$ (respectively $ \mathcal{F}^{(-)}_{\textrm{NS}}$) means the subspace of the Neveu-Schwarz sector with even (respectively odd) fermion parity. The non-Abelian sector on the other hand involves the Ramond sector, and shifted $\mathrm{U}(1)$ charges, reflecting the fractional charge $q_\sigma = 1/2$ of the non-Abelian anyon $\sigma$:
\begin{align}
    \label{eq:H1}
    \mathcal{H}_\sigma & =  \mathcal{F}_{\textrm{R}} \otimes \left( \bigoplus_{q \in \mathbb{Z} + 1/2} \mathcal{F}_{q} \right) 
\end{align}

\section{Li-Haldane at leading order}
\label{app:modular}

At leading order [Eq.~\eqref{eq:asymmetriclihaldanesubbed}], the entanglement Hamiltonian is quadratic, allowing for an explicit computation of the charged moments. Indeed the charged moments can be expressed in terms of ``partition functions" of the form 
\begin{align}
    F^{(\delta)}_a (\alpha | \tau_f,\tau_b) = \textrm{Tr}_{\mathcal{H}^{(\delta)}_a }  \left( e^{i\alpha J_0} q_f^{\left( L^{(f)}_0 - c_f/24 \right)}  q_b^{\left( L^{(b)}_0 - c_b/24 \right)} \right),  
\end{align}
where $q_f = e^{i 2\pi \tau_f}$ and $q_b = e^{i 2\pi \tau_b}$. The spectral flow parameter $\delta$ shifts the $\mathrm{U}(1)$ charges in the topological sectors as follows: 
\begin{align}
    \label{eq:h1spectralflow}
    \mathcal{H}^{(\delta)}_{\mathds{1}} & =  \mathcal{F}^{(+)}_{\textrm{NS}} \otimes \left( \bigoplus_{q \in 2\mathbb{Z} + \delta} \mathcal{F}_{q} \right) \, \oplus \, \mathcal{F}^{(-)}_{\textrm{NS}} \otimes \left( \bigoplus_{q \in 2\mathbb{Z} + 1+ \delta} \mathcal{F}_{q} \right),
\end{align}
while $\mathcal{H}^{(\delta)}_{\psi}  = \mathcal{H}^{(\delta+1)}_{\mathds{1}}$, and 
\begin{align}
    \mathcal{H}^{(\delta)}_{\sigma} & =  \mathcal{F}_{\textrm{R}} \otimes \left( \bigoplus_{q \in \mathbb{Z} + 1/2 + \delta} \mathcal{F}_{q} \right) \,.
\end{align}

The charged moments can be computed exactly in terms of generalized theta functions, 
defined according to the conventions outlined in Appendix~\ref{app:theta}. In the non-Abelian sector, the result takes the form
\begin{align}
    F^{(\delta)}_\sigma (\alpha | \tau_f,\tau_b) =  \sqrt{\frac{\theta_2(\tau_f)}{2 \eta(\tau_f)}} \frac{1}{\eta(\tau_b)}  \vartheta \left[ \begin{array}{c} \delta  + \frac{1}{2}\\ 0 \end{array} \right] \left( \left. \frac{\alpha}{2\pi} \right| \tau_b \right),
\end{align}
while in the vacuum sector, the expression becomes
\begin{align}
    F^{(\delta)}_{\mathds{1}} (\alpha &| \tau_f,\tau_b)  =  \frac{1}{2}\sqrt{\frac{\theta_3(\tau_f)}{ \eta(\tau_f)}} \frac{1}{\eta(\tau_b)}  \vartheta \left[ \begin{array}{c} \delta  \\ 0 \end{array} \right] \left( \left. \frac{\alpha}{2\pi} \right| \tau_b \right) \nonumber \\
    & + \frac{e^{-i\pi \delta}}{2}\sqrt{\frac{\theta_4(\tau_f)}{ \eta(\tau_f)}} \frac{1}{\eta(\tau_b)}  \vartheta \left[ \begin{array}{c} \delta  \\ \frac{1}{2} \end{array} \right] \left( \left. \frac{\alpha}{2\pi} \right| \tau_b \right) 
\end{align}
The expression for the $\psi$ sector is related by a shift in $\delta$, specifically
$F^{(\delta)}_{\psi} (\alpha | \tau_f,\tau_b) = F^{(\delta+1)}_{\mathds{1}} (\alpha | \tau_f,\tau_b)$, 
that is to say 
\begin{align}
    F^{(\delta)}_{\psi} (\alpha  & | \tau_f,\tau_b)  =  \frac{1}{2}\sqrt{\frac{\theta_3(\tau_f)}{ \eta(\tau_f)}} \frac{1}{\eta(\tau_b)}  \vartheta \left[ \begin{array}{c} \delta  \\ 0 \end{array} \right] \left( \left. \frac{\alpha}{2\pi} \right| \tau_b \right) \nonumber \\
    & - \frac{e^{-i\pi \delta}}{2}\sqrt{\frac{\theta_4(\tau_f)}{ \eta(\tau_f)}} \frac{1}{\eta(\tau_b)}  \vartheta \left[ \begin{array}{c} \delta  \\ \frac{1}{2} \end{array} \right] \left( \left. \frac{\alpha}{2\pi} \right| \tau_b \right) 
\end{align}

\subsection{Topological entanglement entropy} 
\label{app:modular_TEE}

As a preliminary consistency check, we examine the large-$L$ behavior of the entanglement entropy to confirm that the mismatch between the neutral and charged velocities does not alter the topological entanglement entropy. From the expressions above, the asymptotic behavior as $L\to \infty$ is readily extracted. 
In the Abelian sectors $a = \mathds{1}, \psi$, we find:
\begin{align}
    \log F^{(\delta)}_a \left( 0 \left| i \frac{v_f}{L},  i \frac{v_b}{L} \right) \right. \sim   \alpha_0 L     - \log 2 + \cdots ,
\end{align}
while in the non-Abelian sector 
\begin{align}
    \log F^{(\delta)}_{\sigma} \left( 0 \left| i \frac{v_f}{L},  i \frac{v_b}{L} \right) \right.   \sim  \alpha_0 L    - \log \sqrt{2} + \cdots .
\end{align}
Here, the non-universal coefficient $\alpha_0$ is determined by the velocities and the central charges of the respective sectors: 
\begin{align}
    \label{eq:alpha}
    \alpha_0 = \frac{\pi}{12}\left( \frac{c_f}{v_f} + \frac{c_b}{v_b} \right).
\end{align}

This leads to the following large-$L$ behavior for the $n^\textrm{th}$ R\'{e}nyi entropy in a topological sector $a$: 
\begin{align}
    \label{eq:renyientropy}
    S_n \sim \frac{n+1}{n} \alpha_0 L - \gamma_a + \cdots ,
\end{align}
with the expected topological entanglement entropies $\gamma_a = \log 2$ for the Abelian sectors and $\gamma_{\sigma} = \log \sqrt{2}$ for the non-Abelian sector.

Thus, while the velocity mismatch modifies the non-universal prefactor $\alpha_0$, it leaves the universal contribution, the topological entanglement entropy, unchanged.

\subsection{FCS}
\label{app:modular_FCS}

\subsubsection{FCS in the $\sigma$ sector}

In the non-Abelian sector $\sigma$ the FCS is described by the discrete Gaussian distribution 
\begin{align}
    \label{eq:app_discrete_gaussian}
    p_q = \frac{1}{Z_{\sigma}^{(\Phi)}} e^{- \frac{q^2}{2\sigma^2}}, \qquad \sigma^2 = \frac{L}{2\pi v_b}, \qquad q \in \mathbb{Z} + \frac{1}{2} + \Phi  \,,
\end{align}
where the normalization factor is given by
\begin{align}
    Z_{\sigma}^{(\Phi)} = \sum_{q \in \mathbb{Z} + \frac{1}{2} + \Phi} e^{- \frac{q^2}{2\sigma^2}} =   \vartheta \left[ \begin{array}{c} \Phi  + \frac{1}{2}\\ 0 \end{array} \right] \left( \left. 0 \right| \tau_b \right),
\end{align}
which at large $L$ behaves asymptotically as 
\begin{align}
    Z_{\sigma}^{(\Phi)} \sim  \sqrt{\frac{L}{v_b}} = \sqrt{2\pi \sigma^2}.
\end{align}
The cumulant generating function does not depend on the neutral velocity:
\begin{align}
    \log \left\langle e^{i\alpha J_0} \right\rangle_{\sigma} & =   \log \frac{F^{(\Phi)}_{\sigma} \left( \frac{\alpha}{2\pi} \left| i \frac{v_f}{L},  i \frac{v_b}{L} \right) \right.}{F^{(\Phi)}_{\sigma} \left( 0 \left| i \frac{v_f}{L},  i \frac{v_b}{L} \right) \right. } \\
    & = \log \frac{\vartheta \left[ \begin{array}{c} \Phi  + \frac{1}{2}\\ 0 \end{array} \right] \left( \left. \frac{\alpha}{2\pi} \right| \tau_b \right)}{\vartheta \left[ \begin{array}{c} \Phi  + \frac{1}{2}\\ 0 \end{array} \right] \left( \left. 0 \right| \tau_b \right)} ,
\end{align}
and it behaves at large $L$ as 
\begin{align}
    \log \left\langle e^{i\alpha J_0} \right\rangle_{\sigma} \sim - \frac{L \alpha^2}{4\pi v_b} +O\left( e^{- \frac{\pi L}{v_b}}\right) .
\end{align}
In the large $L$ regime the variance obeys an area law, in the sense that it grows linearly in $L$, while all the higher cumulants vanish exponentially: 
\begin{align}
    \label{eq:kappa2n}
    \kappa_{2n} \sim (-1)^{n}  2 \cos( 2\pi \Phi ) \left(  \frac{L}{v_b} \right)^{2n} e^{- \frac{\pi L}{v_b}}, \quad n \geq 2
\end{align}
and
\begin{align}
    \kappa_{2n+1} \sim (-1)^{n-1}  2 \sin( 2\pi \Phi ) \left(  \frac{L}{v_b} \right)^{2n+1} e^{- \frac{\pi L}{v_b}}.
\end{align}

\subsubsection{FCS in the Abelian sectors}
\label{app:fcsabelian}

In the Abelian sectors one has instead, for the vacuum sector,
\begin{align} 
    p_q = \frac{1}{Z^{(\Phi)}_{\mathds{1}}} \left\{ \begin{array}{ccc} \left( 1 + \sqrt{\frac{\theta_4(\tau_f)}{\theta_3(\tau_f)}} \right) e^{- \frac{q^2}{2\sigma^2}}  & \textrm{ for } & q \in 2\mathbb{Z} + \Phi  \\ \left( 1 -\sqrt{\frac{\theta_4(\tau_f)}{\theta_3(\tau_f)}} \right) e^{- \frac{q^2}{2\sigma^2}}  & \textrm{ for } & q \in 2\mathbb{Z} + \Phi +1 \end{array} \right. ,
\end{align}
whereas, for the $\psi$ sector, 
\begin{align} 
    p_q = \frac{1}{Z^{(\Phi)}_{\psi}} \left\{ \begin{array}{ccc} \left( 1 - \sqrt{\frac{\theta_4(\tau_f)}{\theta_3(\tau_f)}} \right) e^{- \frac{q^2}{2\sigma^2}}  & \textrm{ for } & q \in 2\mathbb{Z} + \Phi  \\ \left( 1 + \sqrt{\frac{\theta_4(\tau_f)}{\theta_3(\tau_f)}} \right) e^{- \frac{q^2}{2\sigma^2}}  & \textrm{ for } & q \in 2\mathbb{Z} + \Phi +1 \end{array} \right. ,
\end{align}
where $Z^{(\Phi)}_{\mathds{1}}$ and $Z^{(\Phi)}_{\psi}$ are the corresponding normalization factors.

In the vacuum sector the probability to have even fermion parity is then given by
\begin{align}
    p_{\mathrm{even}} =  \frac{1}{Z^{(\Phi)}_{\mathds{1}}} \left( 1 + \sqrt{\frac{\theta_4(\tau_f)}{\theta_3(\tau_f)}} \right)  \vartheta \left[ \begin{array}{c}  \frac{\Phi}{2}\\ 0 \end{array} \right] \left( \left. 0 \right| 4\tau_b \right)
\end{align} 
while the probability to have odd fermion parity is 
\begin{align}
    p_{\mathrm{odd}} =  \frac{1}{Z^{(\Phi)}_{\mathds{1}}} \left( 1 - \sqrt{\frac{\theta_4(\tau_f)}{\theta_3(\tau_f)}} \right)  \vartheta \left[ \begin{array}{c}  \frac{\Phi+1}{2}\\ 0 \end{array} \right] \left( \left. 0 \right| 4\tau_b \right),
\end{align} 
whereas in the $\psi$ sector, the probability to have even fermion parity is given by
\begin{align}
    \label{eq:peven_app}
    p_{\mathrm{even}} =  \frac{1}{Z^{(\Phi)}_{\psi}} \left( 1 + \sqrt{\frac{\theta_4(\tau_f)}{\theta_3(\tau_f)}} \right)  \vartheta \left[ \begin{array}{c}  \frac{\Phi+1}{2}\\ 0 \end{array} \right] \left( \left. 0 \right| 4\tau_b \right)
\end{align} 
and that of odd fermion parity is 
\begin{align}
    \label{eq:podd_app}
    p_{\mathrm{odd}} =  \frac{1}{Z^{(\Phi)}_{\psi}} \left( 1 - \sqrt{\frac{\theta_4(\tau_f)}{\theta_3(\tau_f)}} \right)  \vartheta \left[ \begin{array}{c}  \frac{\Phi}{2}\\ 0 \end{array} \right] \left( \left. 0 \right| 4\tau_b \right).
\end{align} 

Asymptotically at large $L$, we find that the parity imbalance is
\begin{align}
    \label{eq:pimbalance_app}
    p_{\mathrm{even}} - p_{\mathrm{odd}} \sim \sqrt{2} e^{ - \frac{\pi L}{8 v_f}} \pm 2 \cos (\pi \Phi) e^{- \frac{\pi L}{4 v_b}} 
\end{align} 
for the vacuum and $\psi$ sectors, respectively.

\subsection{Symmetry-resolved entropies}

In the non-Abelian sector we have strict equipartition within the leading-order approximation of the modular Hamiltonian. Indeed 
\begin{align}
    \Tr\left[\rho_A(q)^n\right] = 2^{(n-1)/2}\sqrt{\frac{\theta_2(n\tau_f) \eta(\tau_f)^n}{ \theta_2(\tau_f)^n  \eta(b\tau_f)}} \frac{\eta(\tau_b)^n}{\eta(n\tau_b)}
\end{align}
does not depend on $q$. In the Abelian sectors, the symmetry-resolved entanglement entropy depends only on the fermion parity, even or odd:
\begin{align}
    \Tr\left[\rho_A(q)^n\right] = 2^{n-1} \frac{\left( \sqrt{\frac{\theta_3(n\tau_f)}{\eta(n\tau_f)}} \pm  \sqrt{\frac{\theta_4(n\tau_f)}{\eta(n\tau_f)}} \right)}{\left( \sqrt{\frac{\theta_3(\tau_f)}{\eta(\tau_f)}} \pm  \sqrt{\frac{\theta_4(\tau_f)}{\eta(\tau_f)}} \right)^n} \frac{\eta(\tau_b)^n}{\eta(n \tau_b)},
\end{align}
where the sign $\pm$ corresponds to fermion parity ($+$ for even, $-$ for odd). At large $L$ equipartition is recovered exponentially fast since
\begin{align}
    S_n(q) & = \frac{1}{1-n} \log \left( 1 \pm \sqrt{\frac{\theta_4(n\tau_f)}{\theta_3(n\tau_f)}} \right)\left( 1 \pm \sqrt{\frac{\theta_4(\tau_f)}{\theta_3(\tau_f)}} \right)^{-n} 
\end{align}
up to an additive constant that does not depend on fermion parity. 
In all sectors the symmetry-resolved entropies obey the following asymptotic behavior  
\begin{align}
    S_n(q) \sim \frac{n+1}{n} \alpha_0 L - \gamma_a  - \left(  \frac{1}{2} \log \frac{L}{v_b} + \frac{\log n}{2(n-1)} \right) ,
\end{align}
where $\alpha_0$, which depends on both the bosonic and fermionic velocities $v_b$ and $v_f$, is given by Eq.~\eqref{eq:alpha}, and the TEE $\gamma_a$ is given by $\gamma_a = \log 2$ in the Abelian sectors and $\gamma_a = \log \sqrt{2}$ in the $\sigma$ sector, as in Eq.~\eqref{eq:renyientropy}. 
In particular, we recover the decomposition [Eq.~\eqref{eq:total_entropy_sum_rule}] of the von Neumann entropy
\begin{align}
    S_1(q) = S_1 + \sum_q p_q \log p_q ,
\end{align}
where the total von Neumann entropy $S_1$ scales as
\begin{align}
    S_1 \sim 2 \alpha_0 L - \gamma_a + \cdots ,
\end{align}
and the Shannon entropy of charge fluctuation $S^{\textrm{number}} = - \sum_q p_q \log p_q $ behaves, for the discrete Gaussian distribution of Eq.~\eqref{eq:app_discrete_gaussian}, as 
\begin{align}
    S^{\textrm{number}}  \sim  \frac{1}{2} \log 2\pi \sigma^2  + \frac{1}{2}  \sim \frac{1}{2} \log \frac{L}{v_b} + \frac{1}{2} \,.
\end{align}
In contrast to the $\alpha_0$-dependent $S_1$, $S^{\textrm{number}}$ here depends only on the bosonic velocity $v_b$, which makes sense, as $v_b$ is the velocity of the charged mode.

\section{Theta Functions and Conventions}
\label{app:theta}

This appendix serves to fix the notations and conventions used throughout the text for Jacobi theta functions, the Dedekind eta function, and generalized theta functions. The four Jacobi theta functions $\theta_i(z|\tau)$ are defined via the following Fourier series:
\begin{align}
    \theta_1(z|\tau) & = -i \sum_{r \in \mathbb{Z} + 1/2} (-1)^{r-1/2} e^{2\pi i r z} e^{i \pi \tau r^2} \\
    \theta_2(z|\tau) & = \sum_{r \in \mathbb{Z} + 1/2}  e^{2\pi i r z} e^{i \pi \tau r^2} \\
    \theta_3(z|\tau) & = \sum_{n \in \mathbb{Z}}  e^{2\pi i n z} e^{i \pi \tau n^2} \\
    \theta_4(z|\tau) & = \sum_{n \in \mathbb{Z}}  (-1)^n e^{2\pi i n z} e^{i \pi \tau n^2} 
\end{align}
Here, $z$ is a complex variable, and $\tau$ lies in the upper half-plane, $\mathrm{Im}(\tau) > 0$. We use the shortened notation $\theta_i(\tau)$ for $\theta_i(0|\tau)$. These special functions play a central role in conformal field theory and exhibit well-defined modular transformation properties. The Dedekind eta function, another key modular object, is given by
\begin{equation}
    \eta(\tau) = e^{\frac{i \pi \tau}{12}} \prod_{n = 1}^{\infty} \left(1 - e^{2i \pi \tau n} \right) \, .
\end{equation}
It often appears in modular-invariant combinations and provides a canonical normalization for characters and partition functions. A more general form of the theta function, which includes the Jacobi functions as special cases, is the theta function with characteristics:
\begin{align}
    \vartheta \left[ \begin{array}{c} a \\ b \end{array} \right] (z | \tau) & = \sum_{n \in \mathbb{Z}} e^{\pi i \tau (n+a)^2} e^{2\pi i (n+a)(z+b)}. 
\end{align}
Here, $a, b \in \mathbb{R}$ are real characteristics. Varying these parameters recovers the standard Jacobi theta functions. This generalized form is particularly useful in contexts involving twisted boundary conditions or modular transformations.

\section{More on the synthetic entanglement spectra}
\label{app:syntheticES}

\subsection{Calculation of the corrections to the entanglement spectrum}

To compute the synthetic entanglement spectrum itself, we need to diagonalize the linear combination Eq.~\eqref{eq:lincombhent}. We begin by generating a basis for all of the descendant states of each charge primary state for all of the levels above the primary state in the CFT Hilbert space that we consider. 
For consistency with the MPS data, we cut off the number of descendant levels based on their conformal dimension in each charge sector. In particular, we take descendant levels up to a conformal dimension above the primary state of $P_{\max} = 10$ (as opposed to $P_{\max} = 16$ for the MPS, see Appendix \ref{app:truncation}). Thus the dimension of the CFT Hilbert space accounted for in the synthetic entanglement spectrum for a given charge sector is at most 643.
We can write down these descendant states in terms of the modes $J_{-n}$ and $\psi_{-m}$ of $J(y)$ and $\psi(y)$, which can act like raising operators on the primary states to build up the full Hilbert space. More explicitly, the basis for these descendant states is spanned by $\ket{\chi_{f,q,\{n_i\},\{m_j\}}}$ with  
\begin{equation}
	\label{eq:descendantstate}
	\ket{\chi_{f,q,\{n_i\},\{m_j\}}} = J_{-n_1} \cdots J_{-n_\ell} \psi_{-m_1}\cdots \psi_{-m_r}\ket{f}\otimes\ket{q},
\end{equation}
where $\ket{f}$ is either an NS or R fermionic primary state (see Appendix \ref{app:freefields}), $\ket{q}$ is the [$\mathrm{U}(1)$] Kac-Moody primary state with charge $q$, $n_1, \ldots, n_\ell \in \mathbb{Z}_+$, and $m_1, \ldots, m_r \in \mathbb{Z}_+$ if $\ket{f}\otimes\ket{q} \in \mathcal{H}_\sigma$ or $m_1, \ldots, m_r \in\mathbb{Z}_+ + 1/2$ otherwise.
$\ket{\chi_{f,q,\{n_i\},\{m_j\}}}$ is a descendant state at level $K_{\chi_{f,q,\{n_i\},\{m_j\}}} = \sum_i n_i + \sum_j m_j$ in the fermionic and charge sector of descendants of the state $\ket{f}\otimes\ket{q}$, and it will have conformal dimension $\Delta_{\chi_{f,q,\{n_i\},\{m_j\}}} = K_{\chi_{f,q,\{n_i\},\{m_j\}}}  + q^2/2 + h_{l(f,q)}$, where $h_{l(f,q)}$ refers to the conformal dimension of the WZW primary state from Eq.~\eqref{eq:hlconformaldim}, with $l(f,q) = 1$ if $\ket{f}\otimes\ket{q} \in \mathcal{H}_\sigma$ or $l(f,q) =0$ if $\ket{f}\otimes\ket{q} \not\in \mathcal{H}_\sigma$. 
As mentioned previously, the basis is truncated based on the conformal dimension above the (WZW) primary state, so we keep only states such that $P_{\chi_{f,q,\{n_i\},\{m_j\}}} \equiv \Delta_{\chi_{f,q,\{n_i\},\{m_j\}}} - h_{l(f,q)} \leq P_{\max}$.

With the descendant state basis in terms of the $J_{-n}$ and $\psi_{-m}$, we can likewise write down the zero modes $V_i$, which correspond by Eqs.~\eqref{eq:hiintegral} and \eqref{eq:hiscaling} to the zero modes of the operators in Table \ref{table:integerevenchargeoperators}, in terms of the $J_{-n}$ and $\psi_{-m}$. These forms for the $V_i$ are enumerated in Table \ref{table:modereps}. This allows direct computation of the matrix elements of the $V_i$ in the basis of $\ket{\chi_{f,q,\{n_i\},\{m_j\}}}$. The linear combination of these matrices as in Eq.~\eqref{eq:lincombhent} can then be diagonalized, in each charge primary sector, to give an un-normalized entanglement spectrum, which can then be normalized by properly shifting it up or down (corresponding to the correct normalization prefactor for the reduced density matrix of the full topological sector $a$, $\rho_{A,a} \propto e^{-H_{A,a}}$).

\begin{table*}[hbt]
	\centering
	\begin{tabular}{c | c | c | c | c}
		$i$ & $\Delta_i$ & $\phi_i(y)$ & \multicolumn{2}{c}{$V_i = \left(\frac{L}{\pi}\right)^{\Delta_i-1} \int_0^L \phi_i(y) dy$} \\
		\hline
		\hline
		0 & 2 & $(JJ)(y)$ & \multicolumn{2}{c}{$2\sum_{n \in \mathbb{Z}_+} J_{-n} J_n + J_0^2$} \\
		\hline
		1 & 2 & $-(\psi \partial \psi)(y)$ & $2 \sum_{m \in \mathbb{Z}_+} m \psi_{-m}\psi_m$ & $2 \sum_{m\in\mathbb{Z}_+ -1/2} m \psi_{-m}\psi_m$  \\
		\hline
		2 & 4 & $(\partial J\partial J)(y)$ & \multicolumn{2}{c}{$2\sum_{n \in \mathbb{Z}_+} n^2 J_{-n} J_n$}  \\
		\hline
		3 & 4 & $-(\partial \psi \partial^2 \psi)(y)$ & $2 \sum_{m\in\mathbb{Z}_+} m^3 \psi_{-m}\psi_m$ & $2 \sum_{m\in\mathbb{Z}_+-1/2} m^3 \psi_{-m}\psi_m$ \\
		\hline
        4 & 4 & $((JJ)(JJ))(y)$ & \multicolumn{2}{c}{$2\sum_{n\in\mathbb{Z}_+, i,j \in \mathbb{Z}} :J_{-n-j}J_j::J_{n-i} J_i: + \sum_{i,j \in \mathbb{Z}} :J_{-i} J_i: :J_{-j}J_j:$}\\
		\hline
	    5 & 4 & $-((JJ)(\psi \partial \psi))(y)$ & $\sum_{m,n,j \in \mathbb{Z}}m: \psi_{n-m}\psi_m: :J_{-n-j}J_j:$ & $\sum_{n,j \in \mathbb{Z}}\sum_{m \in \mathbb{Z}+1/2} m: \psi_{n-m}\psi_m: :J_{-n-j}J_j:$  \\
		\hline
		\hline
	\end{tabular}
	\caption{Expressions for the locally conserved zero modes $V_i$ in terms of the Fourier modes $J_{-n}$ and $\psi_{-m}$ of the boson $J(y)$ and the Majorana fermion $\psi(y)$, respectively. $\Delta_i$ indicates the conformal dimension of the operator $\phi_i(y)$, which is integrated to give $V_i$. The symbols $::$ indicate the appropriate (bosonic or fermionic) normal ordering by increasing subscripts.}
	\label{table:modereps}
\end{table*} 

\subsection{Detailed description of the fitting procedure}
\label{sec:detailedescriptionoffit}

In the previous subsection, we described how to compute the synthetic entanglement spectrum for a given choice of parameters $g_i$ in Eq.~\eqref{eq:lincombhent}. However, as noted previously, these $g_i$ are non-universal and depend on microscopic details. Thus, they are not known in advance: to build the synthetic entanglement spectrum for a particular topological state, we need to determine what they are for that state. This can be done in a number of ways. One approach is to fit the synthetic entanglement spectrum to the entanglement spectrum of the MPS data. We can write down a fitting function
\begin{equation}
	\label{eq:Rxj}
	R(\{g_i\}) = \sum_j W_j \cdot \left[\xi_{\textrm{MPS},j}-\xi_{\textrm{synthetic},j}\left(\{g_i\}\right)\right]^2,
\end{equation} 
where the $\xi_{\textrm{MPS},j}$ are the levels (called here energies) of the MPS entanglement spectrum, the $\xi_{\textrm{synthetic},j}$ are the energies of the synthetic entanglement spectrum, and the $W_j$ are an associated weighting function defined below. It is not possible to ascertain the true correspondence between the $\xi_{\textrm{MPS},j}$ and the $\xi_{\textrm{synthetic},j}$, so both lists are sorted by energy, within each charge primary sector and momentum, and then the differences in the argument of the quadratic in Eq.~\eqref{eq:Rxj} are calculated between the corresponding elements of the sorted lists. Meanwhile, the $j$th weight $W_j$ is given by
\begin{equation}
	W_j = \frac{1}{N_j}\exp\left(-\xi_{\textrm{MPS},j}\right),
\end{equation}
where $N_j$ is a number equal to the number of states at the same descendant level in the same charge primary sector as the state associated to the $\xi_{\textrm{MPS},j}$ entanglement energy eigenvalue.
Minimizing $R$ as a function of the $g_i$ that determine the entanglement spectrum $\xi_{\textrm{synthetic}}$, we can find the set of $g_i$ that lead to a $\xi_{\textrm{synthetic}}$ of best fit that most closely matches (subject to the weighting $W_j$) the MPS entanglement spectrum $\xi_{\textrm{MPS}}$. The $g_i$ obtained in this way are enumerated in the tables of the next subsection. 

We also benchmark this approach by applying a similar method to compute the synthetic entanglement spectrum for an analytically exact Integer Quantum Hall Effect entanglement spectrum. This test is described in Appendix \ref{app:iqhebenchmark}.

\subsection{Fitting data}
\label{app:fittingdata}

In Tables \ref{table:expsupfitevenchargesdata} and \ref{table:expsupfittwoparameterdata}, we enumerate the parameters $g_i$ of Eq.~\eqref{eq:lincombhent} for the synthetic entanglement spectrum found by the exponentially suppressed fits to the MPS data, as described in the previous subsection. In Table \ref{table:expsupfitevenchargesdata}, these fits are performed with the full set of six integrals of the $\phi_i(y)$ of Table \ref{table:integerevenchargeoperators}, while in Table \ref{table:expsupfittwoparameterdata} we instead try to fit the whole MPS entanglement spectrum with just the integrals of the operators $\phi_i(y)$ of dimension $\Delta_i = 2$: $(JJ)(y)$ and $-(\psi \partial \psi)(y)$. For both tables we include several approaches to the fit with the $\Phi = 1/2$ MPS data. The first column of each table denotes the choice between fitting based on one particular system size ($L = 8$, 10, or 12), and simultaneously fitting all the analyzed system sizes ($L = 8$ through $L = 12$). The second column of each table denotes the choice between fitting either the $\psi$ or $\sigma$ sector on its own, and fitting both sectors simultaneously. (At $\Phi = 1/2$, the vacuum sector contains essentially the same data as the $\psi$ sector, so of the two Abelian sectors, we only consider the $\psi$ sector for these fits.) The resulting sets of $g_i$ are broadly consistent across these different approaches, within each table. Certainly Tables \ref{table:expsupfitevenchargesdata} and \ref{table:expsupfittwoparameterdata} differ substantially, as the quality of the fits in Table \ref{table:expsupfittwoparameterdata} is much worse, as they do not account for any of the $\Delta_i = 4$ integrals of Table \ref{table:modereps} that contribute substantially to shaping the entanglement spectrum. The row indicating the data fit for the $\psi$ sector at $L = 12$ of Table \ref{table:expsupfitevenchargesdata} is shown in bold, as this is the set of fitting parameters used for most of the plots of various quantities displayed in Section \ref{sec:ESfitresults} and Appendix \ref{sec:mpsescomparisons}.

\begin{table}[]
    \centering
    \begin{tabular}{l|l|cccccc}
        \multirow{2}{*}{$L$ fit} & \multirow{2}{*}{Sector} & \multicolumn{6}{c}{Fit coefficients} \\ 
        \cline{3-8} &  & $g_0$ & $g_1$ & $g_2$ & $g_3$ & $g_4$ & $g_5$ \\
        \hline
        \hline
        \multirow{3}{*}{$8$}  & $\psi$  & 2.006 & 1.141 & 2.139 & 1.244 & 0.289 & 1.618 \\
        \cline{2-8}
         & $\sigma$ & 1.938 & 0.838 & 1.954 & 1.849 & 0.320 & 1.958 \\
        \cline{2-8}
         & both  & 1.996 & 1.300 & 2.284 & 0.923 & 0.302 & 1.320 \\
        \hline
        \multirow{3}{*}{$10$} & $\psi$  & 1.887 & 0.892 & 2.001 & 1.604 & 0.329 & 1.969 \\
        \cline{2-8}
         & $\sigma$ & 1.844 & 0.774 & 1.744 & 1.849 & 0.361 & 2.125 \\
        \cline{2-8}
         & both  & 1.885 & 0.980 & 2.018 & 1.393 & 0.341 & 1.696 \\
        \hline 
        \multirow{3}{*}{$12$} & $\psi$  & \textbf{1.819} & \textbf{0.774} & \textbf{1.818} & \textbf{1.807} & \textbf{0.362} & \textbf{2.223} \\
        \cline{2-8}
         & $\sigma$ & 1.799 & 0.719 & 1.646 & 1.933 & 0.388 & 2.311 \\
        \cline{2-8}
         & both  & 1.823 & 0.815 & 1.829 & 1.691 & 0.370 & 2.031 \\
        \hline 
        \multirow{3}{*}{All} & $\psi$  & 1.967 & 1.078 & 2.362 & 1.315 & 0.299 & 1.754 \\
        \cline{2-8}
         & $\sigma$ & 1.905 & 0.823 & 1.971 & 1.835 & 0.332 & 2.089 \\
        \cline{2-8}
         & both  & 1.953 & 1.149 & 2.361 & 1.120 & 0.314 & 1.500 \\
        \hline
    \end{tabular} 
    \caption{The parameters $g_i$ of Eq.~\eqref{eq:lincombhent} for the synthetic entanglement spectrum found by the exponentially suppressed fits to the MPS data at $\Phi = 1/2$. These fits are performed with the full set of six integrals of the $\phi_i(y)$ of Table \ref{table:integerevenchargeoperators}. Several approaches to the fit are included. The first column denotes the choice between fitting based on one particular system size ($L = 8$, 10, or 12), and simultaneously fitting all the analyzed system sizes ($L = 8$ through $L = 12$). The second column denotes the choice between fitting either the $\psi$ or $\sigma$ sector on its own, and fitting both sectors simultaneously. The fit from the $\psi$ sector at $L = 12$, which is used for most of the plots generated from the synthetic entanglement spectrum fits, is highlighted in bold.}
    \label{table:expsupfitevenchargesdata}
\end{table}

\begin{table}[]
    \centering
    \begin{tabular}{l|l|cc}
        \multirow{2}{*}{$L$ fit} & \multirow{2}{*}{Sector} & \multicolumn{2}{c}{Fit coefficients} \\ 
        \cline{3-4} &  & $g_0$ & $g_1$  \\
        \hline
        \hline
        \multirow{3}{*}{$L = 8$}  & $\psi$ & 3.22388 & 2.83382 \\
        \cline{2-4}
         & $\sigma$ & 3.04171 & 2.64434 \\
        \cline{2-4}
         & both  & 3.13744 & 3.02579 \\
        \hline
        \multirow{3}{*}{$L = 10$} & $\psi$  &  2.82089 & 2.53442 \\
        \cline{2-4}
         & $\sigma$ & 2.6862 & 2.33886 \\
        \cline{2-4}
         & both  & 2.75991 & 2.56431 \\
        \hline 
        \multirow{3}{*}{$L = 12$} & $\psi$  & 2.57688 & 2.28082 \\
        \cline{2-4}
         & $\sigma$ & 2.46943 & 2.1088 \\
        \cline{2-4}
         & both  & 2.52956 & 2.25589 \\
        \hline
        \multirow{3}{*}{All} & $\psi$  & 2.93408 & 2.72468 \\
        \cline{2-4}
         & $\sigma$ & 2.79174 & 2.55173 \\
        \cline{2-4}
         & both  & 2.86829 & 2.76663 \\
    \end{tabular} 
    \caption{The parameters $g_i$ of Eq.~\eqref{eq:lincombhent} for the synthetic entanglement spectrum found by the exponentially suppressed fits to the MPS data, in the case where we try to fit the whole MPS entanglement spectrum with solely the integrals of the operators $\phi_i(y)$ of dimension $\Delta_i = 2$: $(JJ)(y)$ and $-(\psi \partial \psi)(y)$. Several approaches to the fit are included. The first column denotes the choice between fitting based on one particular system size ($L = 8$, 10, or 12), and simultaneously fitting all the analyzed system sizes ($L = 8$ through $L = 12$). The second column denotes the choice between fitting either the $\psi$ or $\sigma$ sector on its own, and fitting both sectors simultaneously.}
    \label{table:expsupfittwoparameterdata}
\end{table}

\section{Benchmarking the synthetic entanglement spectrum method with Integer Quantum Hall Effect data}
\label{app:iqhebenchmark}

One way that we can benchmark our approach based on the synthetic entanglement spectrum is to apply it to the analytically understood case of the Integer Quantum Hall Effect. 
In particular, we would like to validate our fitting procedure 
of Appendix \ref{sec:detailedescriptionoffit} to the $\nu = 1$ IQHE case. We present some of the successful outcomes of this test here. 

To briefly recap, we can build up the entanglement spectrum of the $\nu = 1$ IQHE ground state in the following way. The conformal field theory of the edge is that of a chiral Dirac fermion $\Psi^\dagger(y)$. 
We can decompose this into Fourier modes,
\begin{equation}
	\Psi^{\dagger}(y) \equiv \sum_{m} e^{2\pi i m y/L} c_m^{\dagger}.
\end{equation}
These we can then use to build up the full many-body Hilbert space. At each momentum $k_m = \frac{2\pi m}{L}$, we can write basis states of the form  
\begin{equation}
	\label{eq:diracfermionhilbertspace}
	\ket{\phi_{k_m},i} = c^\dagger_{n_{i,1}} \cdots c^\dagger_{n_{i,N}} \ket{0},
\end{equation}
where $\sum_{j=1}^N n_{i,j} = m$.

We can calculate the entanglement Hamiltonian $H_A$ from knowledge of correlations~\cite{Chung2001,Peschel2003} as
\begin{equation}
	H_A = \sum_m \epsilon(k_m) : c^\dagger_m c_m :,
\end{equation}
where $\epsilon(k) = \log\left[\frac{\textrm{erfc}(-k)}{\textrm{erfc}(k)}\right]$.
Analogously to Eq.~\eqref{eq:lincombhent}, we can then expand $\epsilon(k)$ as a power series to write down $H_A$ in the form
\begin{equation}
	\label{eq:lincombhentiqhe}
	H_A = \sum_{j \geq 0} g_j \sum_{m \in \mathbb{Z}} k_m^{2j+1} :c_m^\dagger c_m:
\end{equation}
where the $g_j$ are coefficients, and the $\sum_{m \in \mathbb{Z}} k_m^{2j+1} :c_m^\dagger c_m:$ terms at each $j$ are the zero modes of dimension $\Delta = 2j+2$ operators $\Psi^\dagger(y)(-i\partial_y)^{2j+1}\Psi(y)$.

We can calculate the matrix elements of these operators by simply applying the Dirac fermion anticommutation relations on each basis state in the Hilbert space [see Eq.~\eqref{eq:diracfermionhilbertspace}]. 
Then we can perform a fit of a truncation of the series in Eq.~\eqref{eq:lincombhentiqhe} in the manner described in Appendix \ref{sec:detailedescriptionoffit} to the exact entanglement spectrum, using the $g_j$ as free parameters.

\begin{figure}
    \centering
    \includegraphics[width=\linewidth]{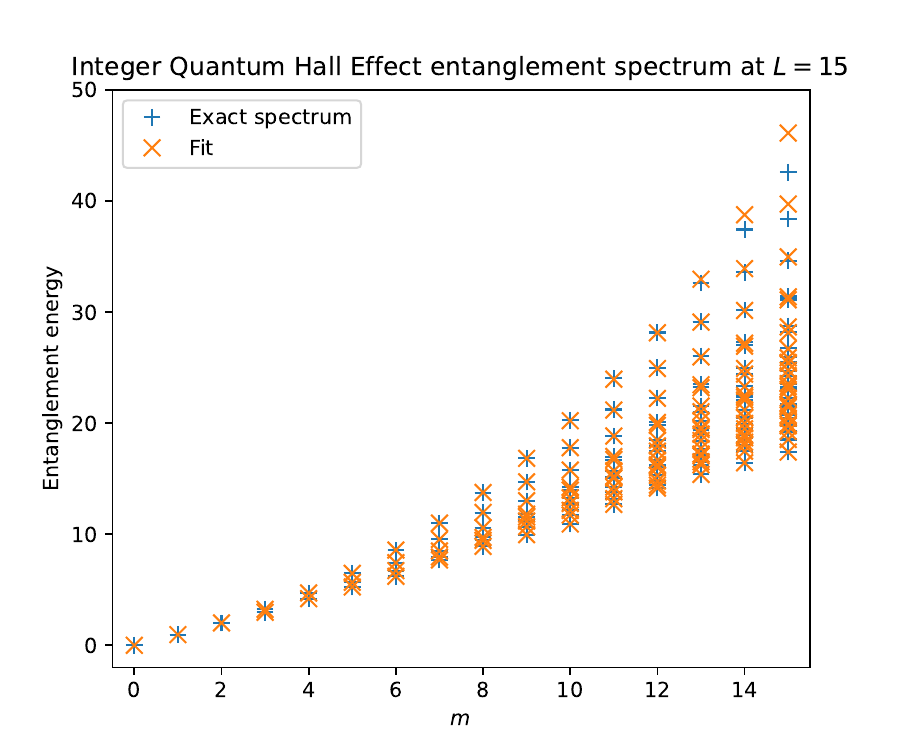}
    \caption{A comparison of the exact integer quantum Hall effect spectrum and the fit performed to the first 13 descendant levels using the first four terms of Eq.~\eqref{eq:lincombhentiqhe}. The parameters used correspond to the $L = 15$ row of Table \ref{table:expsupfitiqhedata}. Each tower of states represents a separate momentum $k_m = \frac{2\pi m}{L}$, labeled by the integer $m$.}
    \label{fig:iqhespectrumcomparison}
\end{figure}

We do this for 13 descendant levels of the entanglement spectrum, at system sizes $L = 5$, 10, 15, and 20, which then give the results for the $g_i$ in the first four data rows of Table \ref{table:expsupfitiqhedata}. But we can calculate the $g_i$ directly from the expanded form of $\epsilon(k)$ as in Eq.~\eqref{eq:lincombhentiqhe}, which obtains the last row of Table \ref{table:expsupfitiqhedata}. The approximation to the actual entanglement spectrum are illustrated in Fig.~\ref{fig:iqhespectrumcomparison}, where we see excellent overlap at the lower descendant levels.

\begin{table}[]
	\centering
	\begin{tabular}{l|cccc}
		\multirow{2}{*}{$L$ fit} & \multicolumn{4}{c}{Fit coefficients} \\ 
		\cline{2-5}  & $g_0$ & $g_1$ & $g_2$ & $g_3$ \\
		\hline
        \hline
		5 &2.25569 & 0.210593 & -0.00482413 & 0.0000668856 \\
		\hline
	    10 & 2.25481 & 0.211673 & -0.00496543 & 0.0000703181 \\
		\hline 
		15 & 2.25496 & 0.211648 & -0.00499315 & 0.0000733395 \\
		\hline 
        20 & 2.25523 & 0.2113 & -0.00488908 & 0.0000655272 \\
		\hline
		\hline
		Power series & 2.25676 & 0.205545 & -0.000418389 & -0.00118025
	\end{tabular} 
	\caption{The parameters $g_i$ of Eq.~\eqref{eq:lincombhent} for the synthetic entanglement spectrum found by the exponentially suppressed fits to the exact integer quantum Hall effect spectrum data. These fits are performed on the first 13 descendant levels with the first four terms of Eq.~\eqref{eq:lincombhentiqhe}, at the respective system sizes given by $L$ in the left column. The final row provides, for comparison, the $g_i$ obtained instead by direct truncation of the power series of $\epsilon(k)$.}
	\label{table:expsupfitiqhedata}
\end{table}

\section{Additional numerical results}
\label{app:additionalnumerical}

\subsection{Truncation analysis of MPS data}
\label{app:truncation}

One important characterization of the MPS data that must be performed is understanding the effect of the truncation in $P_{\max}$. As described in Section \ref{sec:mpsresults} and Appendix \ref{app:syntheticES}, this is the approach of truncating the CFT Hilbert space that is the virtual space of the MPS to include only states of conformal dimension above the primary state $P \leq P_{\max}$. The maximum $P_{\max}$ in the MPS data is $P_{\max} = 16$. In this section, we evaluate the effect of the truncation on the TEE and SREE, using the former to validate for which cylinder perimeters $L$ the MPS data most accurately captures the topological properties of the Moore-Read state. 

We calculate the TEE $\gamma_a(L)$ in each topological sector $a$, for the Moore-Read state at system size $L$, by
\begin{equation}
    \label{eq:teefiniteperimeter}
	-\gamma_a(L) = S_{n,a}(L) - L \frac{\partial S_{n,a}(L)}{\partial L},
\end{equation}
where by $S_{n,a}$ we mean the $n$-R\'{e}nyi entanglement entropy (or von Neumann entropy, for $n =1$) in sector $a$. 
The perimeter being a continuous variable in the MPS approach, we can obtain the derivative $\frac{\partial S_{n,a}(L)}{\partial L}$ by a symmetric difference method.
The topological order of the Moore-Read state is characterized by universal values of $\gamma_a$~\cite{Levin2006,KitaevPreskill2006} equal to $2 \gamma_{\sigma} = \gamma_{\mathds{1}} = \gamma_{\psi} = \log 2$~\cite{Moore1991}
to which the MPS converges for sufficiently large truncation and perimeter~\cite{Estienne2013,Crepel2019c}. 

\begin{figure*}
    \centering
    \subfloat[\label{fig:MooreRead1vacuumTEE1vsL}]{%
      \includegraphics[width=0.6666\columnwidth]{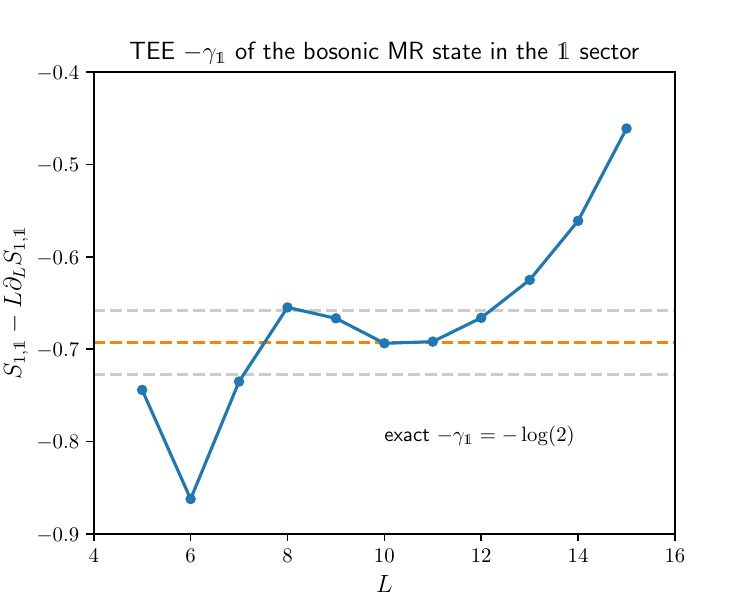}%
    }\hspace*{\fill}%
    \subfloat[\label{fig:MooreRead1sigmaTEE1vsL}]{%
      \includegraphics[width=0.6666\columnwidth]{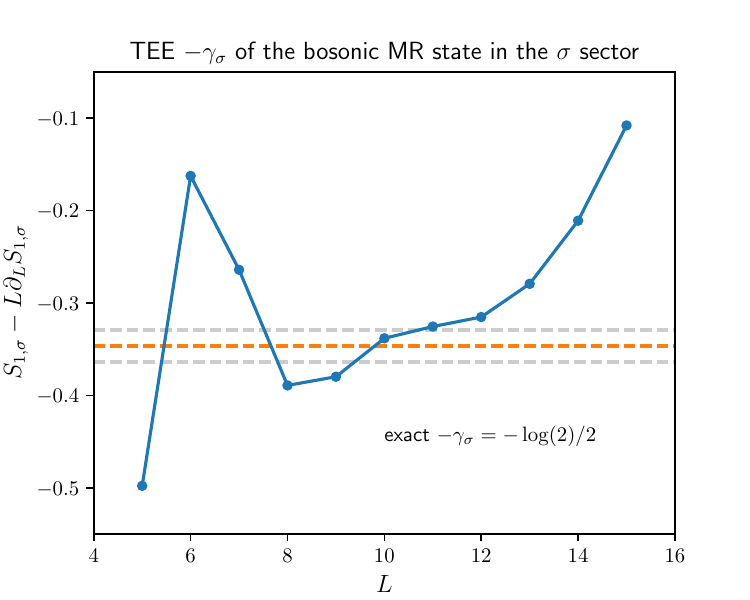}%
    }\hspace*{\fill}%
    \subfloat[\label{fig:MooreRead1psiTEE1vsL}]{%
      \includegraphics[width=0.6666\columnwidth]{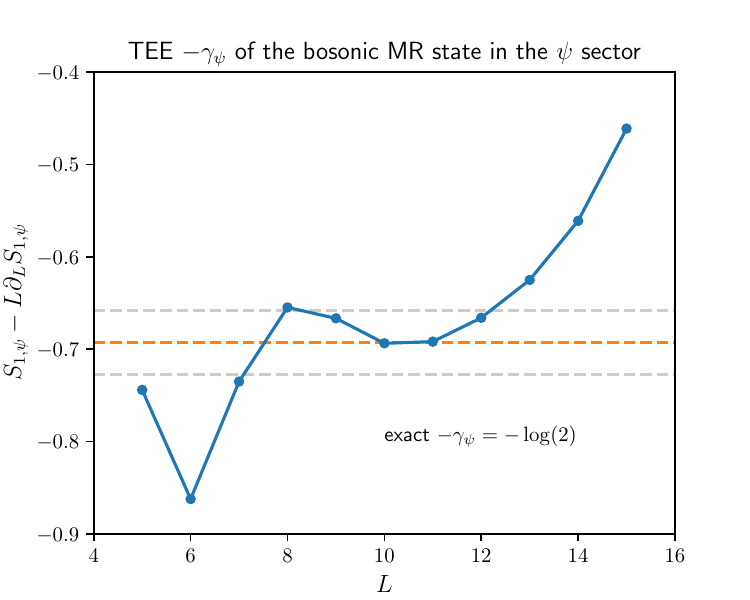}%
    }
    \caption{Von Neumann TEE in the vacuum $\mathds{1}$ (a), $\sigma$ (b), and $\psi$ (c) sectors at flux $\Phi = 1/2$, calculated using finite-difference and Eq.~\ref{eq:teefiniteperimeter}, as a function of cylinder perimeter $L$ for $L = 5,\ldots,15$. The dashed orange line indicates the expected exact value for $-\gamma_{\mathds{1}}$, while the gray dashed lines above and below indicate a $\pm 5\%$ error range.}
    \label{fig:three_graphs_tee}
\end{figure*}

Plots of $-\gamma_a(L)$ vs.~$L$ for the MPS data, calculated from the von Neumann entanglement entropy $S_{1,a}$, are shown in Figs.~\ref{fig:MooreRead1vacuumTEE1vsL}, \ref{fig:MooreRead1sigmaTEE1vsL}, and \ref{fig:MooreRead1psiTEE1vsL}, for the vacuum, $\sigma$, and $\psi$ topological sectors, respectively. These plots are over a range of system sizes studied, from $L = 5$ to $L = 15$. 
The calculated TEE $\gamma_a(L)$ and the expected exact values $\gamma_a$ lie within $\pm 5$ percent (gray dashed lines) in the range $8 \leq L \leq 12$, which constitute the validated system sizes toward which we direct our analysis. 
For smaller perimeters, the MPS data is converged with respect to the truncation parameter but is affected by finite-size effects, whereas for larger perimeters the data is free from finite-size effects but poorly described at the current level of truncation. 

\begin{figure*}
    \centering
    \subfloat[\label{fig:MooreRead1vacuumTEE1vsLtruncated}]{%
      \includegraphics[width=0.6666\columnwidth]{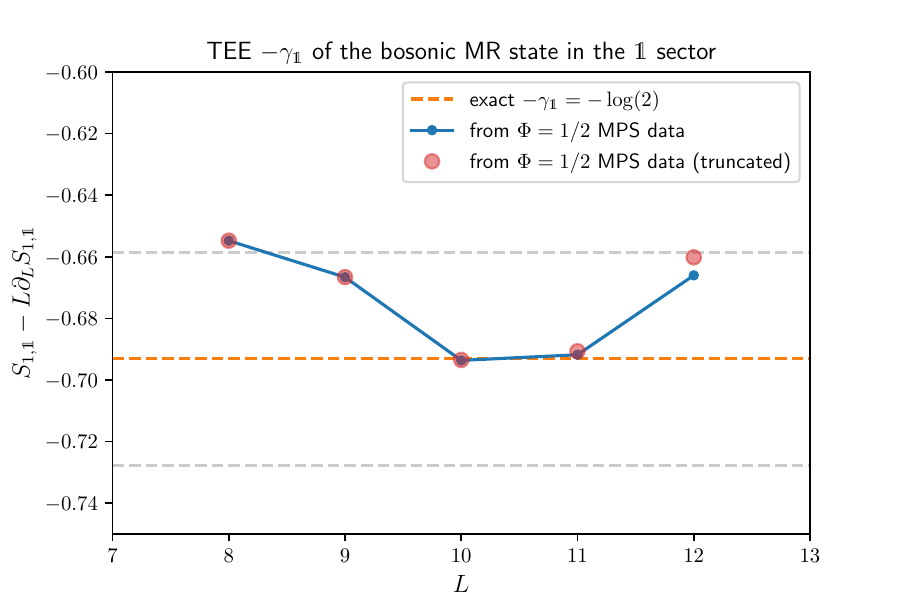}%
    }\hspace*{\fill}%
    \subfloat[\label{fig:MooreRead1sigmaTEE1vsLtruncated}]{%
      \includegraphics[width=0.6666\columnwidth]{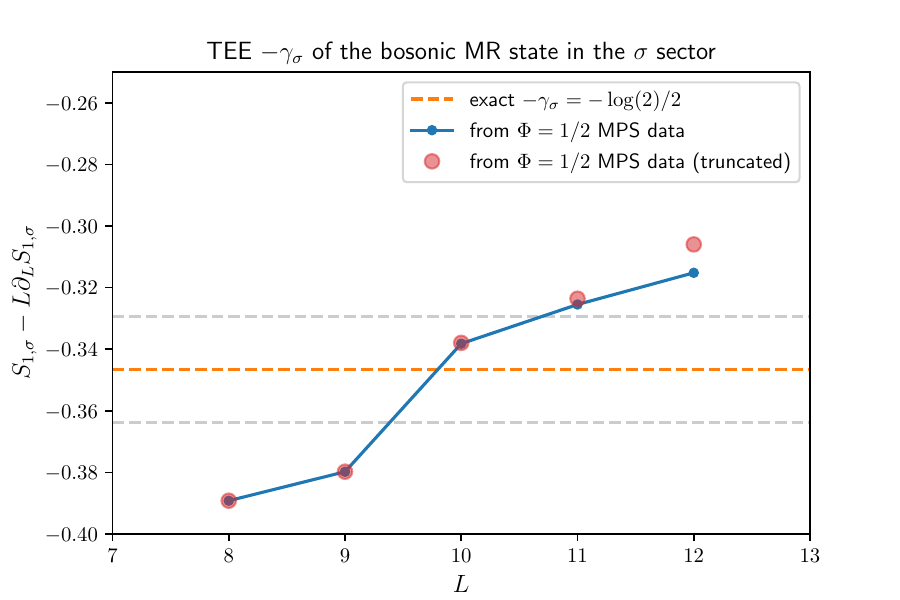}%
    }\hspace*{\fill}%
    \subfloat[\label{fig:MooreRead1psiTEE1vsLtruncated}]{%
      \includegraphics[width=0.6666\columnwidth]{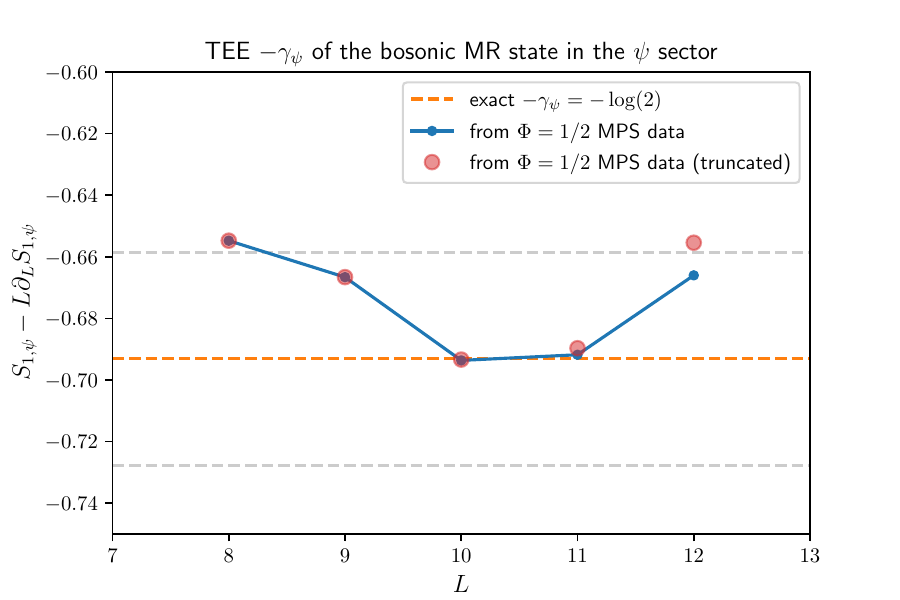}%
    }
    \caption{Comparison of the von Neumann topological entanglement entropy of the bosonic Moore-Read state in the vacuum $\mathds{1}$ sector (a), $\sigma$ sector (b), and $\psi$ sector (c), computed from the $\Phi = 1/2$ MPS data with $P_{\max} = 16$ (blue disks), with that computed from the $\Phi = 1/2$ MPS data truncated to $P_{\max} = 10$ (larger, red disks). The values are shown for cylinder circumference $L = 8,\ldots,12$. In each sector, there is substantial overlap between the MPS data truncated at $P_{\max} = 10$ and the full MPS data, except for a slight deviation at $L = 12$.}
    \label{fig:three_graphs_TEEtruncated}
\end{figure*}

Another truncation question arises when considering the portion of the MPS data to which we perform the fit of the synthetic entanglement spectrum. This is limited to states with conformal dimension above the primary state $P \leq 10$ (i.e., $P_{\max} = 10$) with flux $\Phi = 1/2$. In Fig.~\ref{fig:three_graphs_TEEtruncated}, we confirm that this should not affect the TEE much relative to the $P_{\max} = 16$ of the MPS over the range of $L$ considered, justifying this simplification of the fitting procedure. We do indeed begin to see a modest effect of the truncation at $L = 12$, however, consistent with the greater significance of truncation effects at larger system sizes.

\begin{figure}
    \centering
    \includegraphics[scale=0.6]{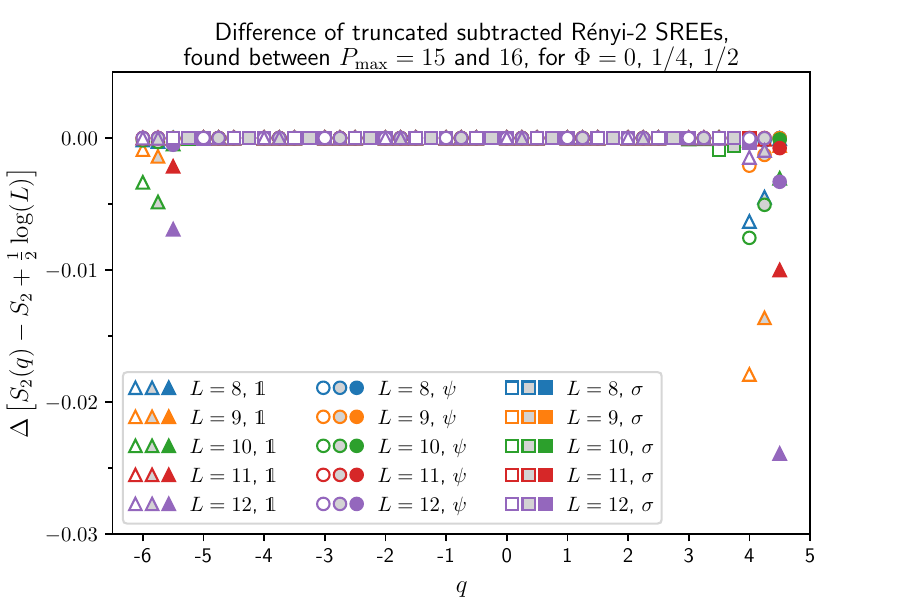}
    \caption{The difference in the subtracted second R\'{e}nyi SREE for the bosonic Moore-Read state calculated between the MPS data truncated at a conformal dimension of $P_{\max} = 16$ and that at $P_{\max} = 15$. This difference is shown for the range of circumferences from $L = 8$ to $L = 12$, and for charges between $q = -6$ and $q = 9/2$, in all three topological sectors, and at fluxes $\Phi = 0, 1/4,$ and $1/2$ (indicated by progressively more filled symbols). The vacuum and $\psi$ sectors are indicated by the triangular and circular markers at half-integer charges, respectively, while the $\sigma$ sector indicated by the square markers at integer charges.}
    \label{fig:MooreRead1comboS2qsubtractedbylihaldaneTruncated1516MPS}
\end{figure}

Finally, we explore the effect of truncation on the SREE by computing the difference of SREEs in all three topological sectors as we reduce the truncation from $P_{\max} = 16$ to $P_{\max} = 15$ for the range of circumferences from $L = 8$ to $L = 12$. This is shown in Fig.~\ref{fig:MooreRead1comboS2qsubtractedbylihaldaneTruncated1516MPS}. We observe that the central part of the charge range is stable with respect to truncating more over the full range of circumferences analyzed. However, at the edges of the chare range, especially at the higher circumferences, the subtracted SREE indicates greater sensitivity to increasing the truncation. These results motivate fitting the quartic fits we fit in Figs.~\ref{fig:MooreRead1comboS2qsubtractedbylihaldaneL8}--\ref{fig:MooreRead1comboS2qsubtractedbylihaldaneL12} solely to data within the charge range $-21/4 \leq q \leq 15/4$, in order to avoid the instabilities seen in the outermost edges of the charge range.

\subsection{Comparison between the $v_b$ and $v_f$ obtained by the two-parameter Gaussian fits to the FCS of the MPS at different $L$}
\label{app:vbvfcomparison}

\begin{figure}
    \centering
    \includegraphics[width=\linewidth]{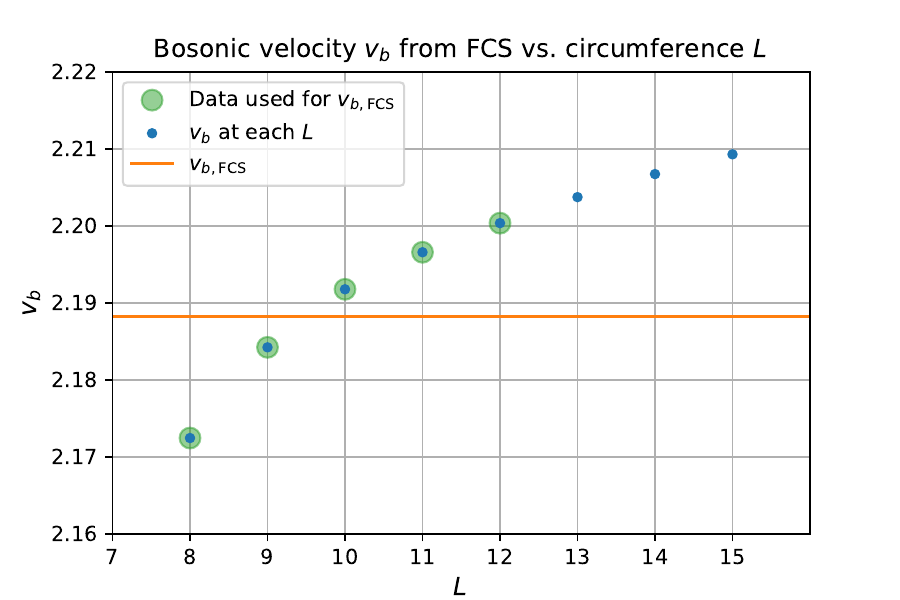}
    \caption{The bosonic velocities $v_b$ calculated from two-parameter Gaussian fits to the FCS in all three topological sectors at fluxes $\Phi = 0$ and $\Phi = 1/2$ and at each cylinder circumference $L$ are depicted by the blue data points. The horizontal orange line illustrates instead the $v_{b,\textrm{fit}} \approx 2.19$ obtained from a simultaneous fit of the same two-parameter Gaussian form to all the data of cylinder circumferences $L = 8$ through $L = 12$ (i.e., the system sizes highlighted in green).}
    \label{fig:MooreRead1averagevbfromfcsMPS}
\end{figure}

\begin{figure}
    \centering
    \includegraphics[width=\linewidth]{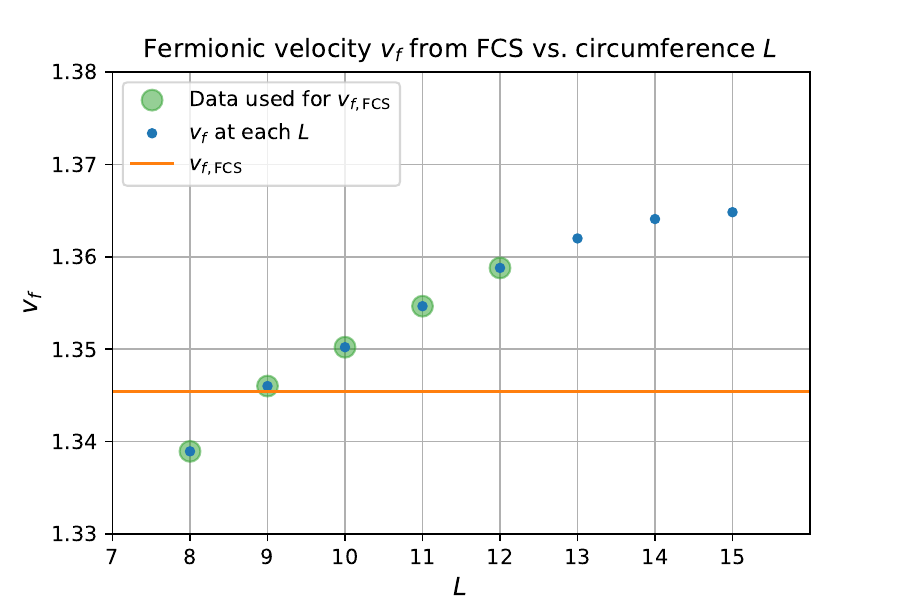}
    \caption{The fermionic velocities $v_f$ calculated from two-parameter Gaussian fits to the FCS in all three topological sectors at fluxes $\Phi = 0$ and $\Phi = 1/2$ and at each cylinder circumference $L$ are depicted by the blue data points. The horizontal orange line illustrates instead the $v_{f,\textrm{fit}} \approx 1.34$ obtained from a simultaneous fit of the same two-parameter Gaussian form to all the data of cylinder circumferences $L = 8$ through $L = 12$ (i.e., the system sizes highlighted in green).}
    \label{fig:MooreRead1averagevffromfcsMPS}
\end{figure}

In Section \ref{sec:mpsresults}, we simultaneously fit the FCS of the MPS in all topological sectors with a single set of Gaussian forms [Eqs.~\eqref{eq:discrete_gaussian} and \eqref{eq:paritygaussian}], depending on only two global parameters $v_{b,\textrm{fit}}$ and $v_{f,\textrm{fit}}$, over the range of cylinder circumferences from $L = 8$ to $L = 12$. We here take an alternative approach, and instead perform the two-parameter Gaussian fit over all topological sectors to the MPS FCS data at each value of $L$ individually, thus obtaining a $v_{b}(L)$ and $v_f(L)$ for each value of $L$. These are plotted in the data points of Figs.~\ref{fig:MooreRead1averagevbfromfcsMPS} and \ref{fig:MooreRead1averagevffromfcsMPS}, respectively, while the lines indicate the values of $v_{b,\textrm{fit}}$ and $v_{f,\textrm{fit}}$ obtained from the simultaneous fit of Gaussians over multiple $L$ performed in the main text. As is apparent, the values at the various $L$ generally cluster pretty close to the values found from that simultaneous Gaussian fit.

\subsection{Cumulant analysis of the MPS FCS data}
\label{app:cumulants}

\begin{figure*}
    \centering
    \subfloat[\label{fig:MooreRead1averagefcsallkappa2MPS}]{%
      \includegraphics[width=0.6666\columnwidth]{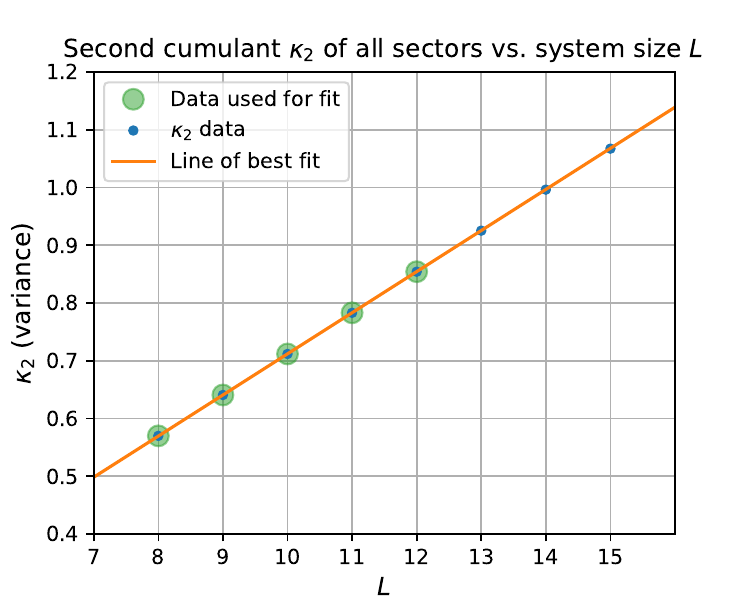}%
    }\hspace*{\fill}%
    \subfloat[\label{fig:MooreRead1averagefcsallkappa4MPS}]{%
      \includegraphics[width=0.6666\columnwidth]{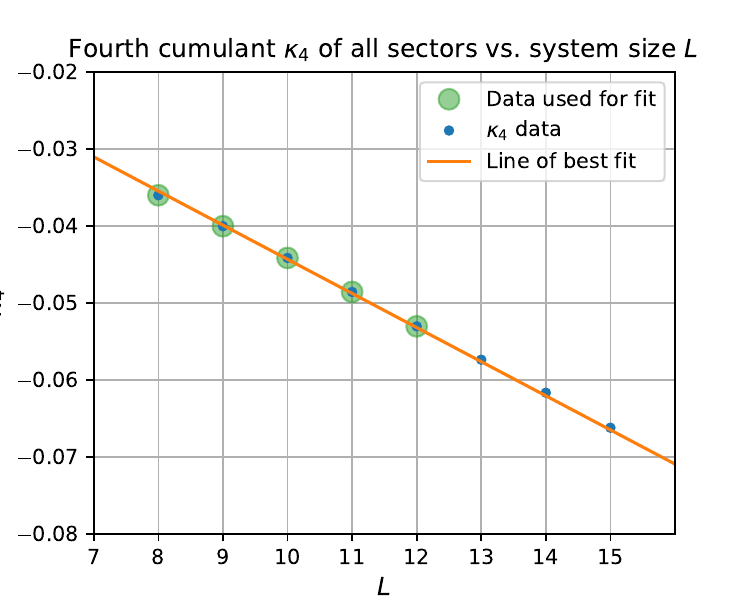}%
    }\hspace*{\fill}%
    \subfloat[\label{fig:MooreRead1averagefcsallkappa6MPS}]{%
      \includegraphics[width=0.6666\columnwidth]{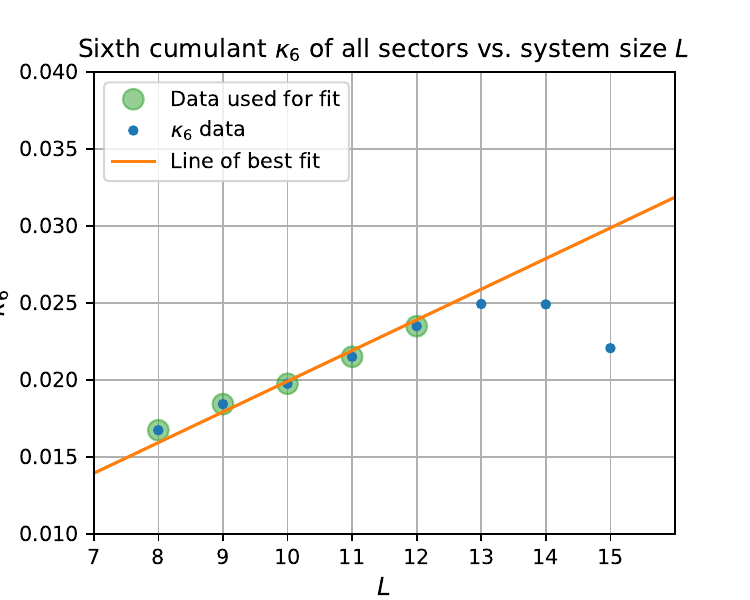}%
    }
    \caption{The second, fourth and sixth cumulants $\kappa_2$ (a) $\kappa_4$ (b) and $\kappa_6$ (c) of the FCS of the $\nu = 1$ Moore-Read state, obtained from the MPS data in all topological sectors and fluxes $\Phi = 0$ and $\Phi = 1/2$, is plotted against the cylinder circumference $L$. The line is the result of a linear fit through the origin, which is performed using only the data of $L = 8$ through $L = 12$, the sizes where the TEE is validated. In (a) the slope of that line (approximately 0.0712) yields an estimate for the bosonic velocity $v_b$ of $v_{b,\kappa_2} \approx 2.24$ using Eq.~\eqref{eq:sigma2L}, keeping in mind that $\kappa_2 = \sigma^2$ (the variance).}
    \label{fig:three_graphs_cumulant}
\end{figure*}

It is also instructive to look at the cumulants of the MPS FCS data, which were considered analytically at the level of the Li-Haldane leading order in Appendix \ref{app:modular_FCS}. We can start with the second cumulant, $\kappa_2$, which is simply the variance of the FCS. This should therefore yield a result close to that obtained via the Gaussian fit to the FCS, into which the bosonic velocity entered via the variance, and for which we found $v_{b,\textrm{FCS}} \approx 2.19$ for the bosonic velocity. Indeed, we obtain a good linear fit, as seen in Fig.~\ref{fig:MooreRead1averagefcsallkappa2MPS}, consistent with Eq.~\eqref{eq:sigma2L}. We can then fit the slope and so obtain the bosonic velocity $v_b$ directly from the second cumulants, and we obtain a value of $v_{b,\kappa_2} \approx 2.24$, close to that obtained directly from the Gaussian fit to the FCS. In Section \ref{sec:consequenceslihaldane}, we noted our expectation of linear (area law) behavior in the higher even cumulants, and we indeed observe this for the case of the fourth and sixth cumulants ($\kappa_4$ and $\kappa_6$, respectively), at least within the range of $L$ where we validated the TEE ($L = 8$ through $L = 12$). This is shown in Figs.~\ref{fig:MooreRead1averagefcsallkappa4MPS} and \ref{fig:MooreRead1averagefcsallkappa6MPS}.

\subsection{Additional comparisons between the MPS and synthetic ES data}
\label{sec:mpsescomparisons}

In this Appendix we consider some additional comparisons between the synthetic ES data and the MPS data: direct comparison of entanglement spectra, the calculation of the symmetry-resolved entanglement entropy, and the FCS for both data sets. 

\begin{figure}
    \centering
    \includegraphics[width=\linewidth]{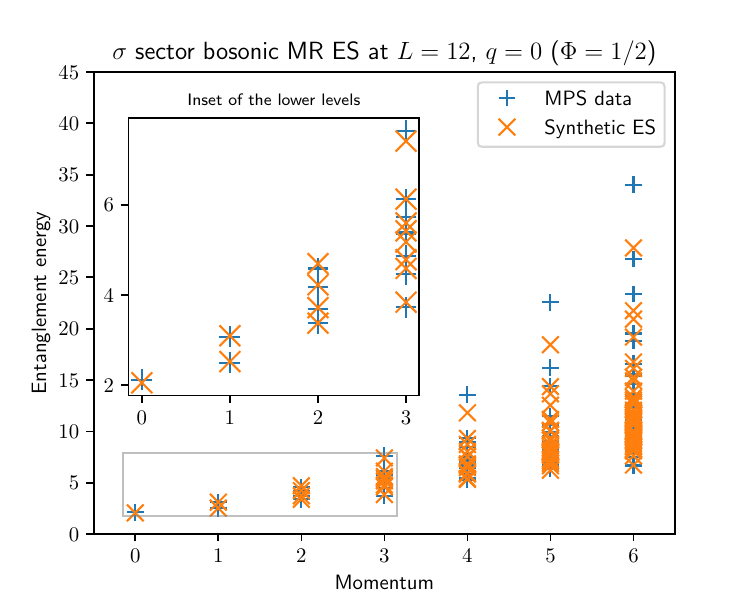}
    \caption{A comparison of the lower levels of the bosonic Moore-Read entanglement spectrum data from the MPS data (blue $+$) and from the synthetic entanglement spectrum fit (orange $\times$), at circumference $L = 12$ and charge $q = 0$ in the $\sigma$ sector (for $\Phi = 1/2$). 
    The data is shown up to an entanglement energy of 45, for the first six descendant levels above the primary state. An inset of the primary state and first three descendant levels reveals the excellent agreement at these lowest levels.}
    \label{fig:mrlowerescomparisonsigma}
\end{figure}

\begin{figure}
    \centering
    \includegraphics[width=\linewidth]{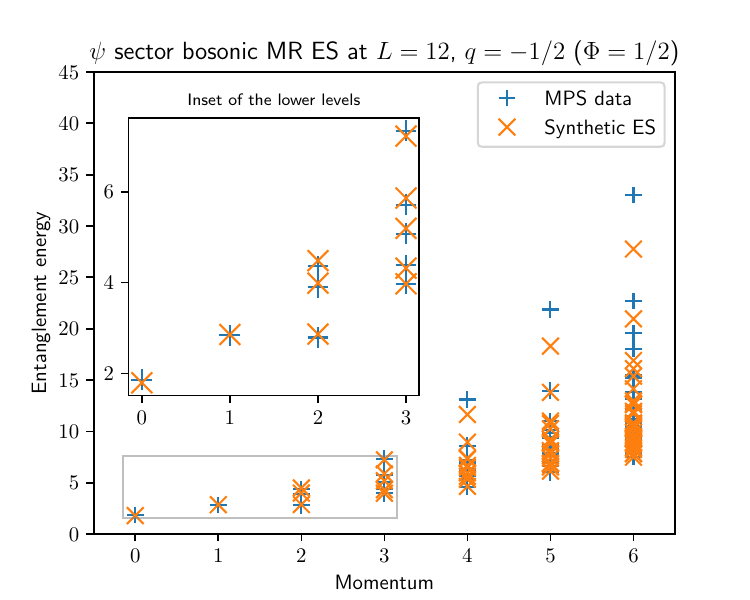}
    \caption{A comparison of the lower levels of the bosonic Moore-Read entanglement spectrum data from the MPS data (blue $+$) and from the synthetic entanglement spectrum fit (orange $\times$), at circumference $L = 12$ and charge $q = -1/2$ in the $\psi$ sector (for $\Phi = 1/2$). 
    The data is shown up to an entanglement energy of 45, for the first six descendant levels above the primary state. An inset of the primary state and first three descendant levels reveals the excellent agreement at these lowest levels.}
    \label{fig:mrlowerescomparisonpsi}
\end{figure}

It is useful to compare the entanglement spectra directly, charge sector by charge sector, to ensure that the synthetic ES data indeed possesses the agreement with the MPS data at the low levels of the ES that it ought to have by construction. Example comparisons are shown for the lower descendant levels at charge $q = 0$, for the $\sigma$ sector at $\Phi = 1/2$, and charge $q = -1/2$, for the $\psi$ sector also at $\Phi = 1/2$, in Figs.~\ref{fig:mrlowerescomparisonsigma} and \ref{fig:mrlowerescomparisonpsi}, respectively. These two sectors not only have the dominant weights, i.e., the largest $p_q$ in the FCS, but also contain the highest number of entanglement levels. As such, they constitute the most stringent test. While the sparser, higher level entanglement energies begin to differ more between the two spectra, we can see relatively good agreement at the lower entanglement energy levels that are most determinative of the spectral properties. 

\begin{figure}
	\centering
	\includegraphics[scale=0.6]{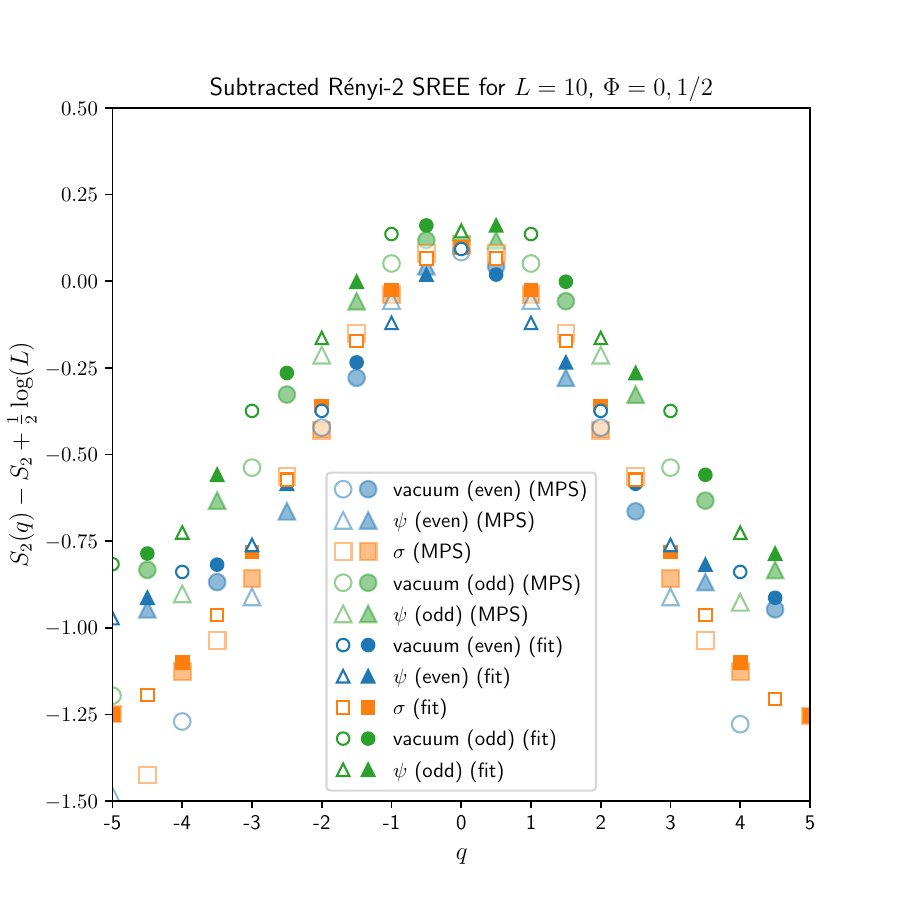}
	\caption{The subtracted symmetry-resolved second R\'{e}nyi entanglement entropy $S_2(q) -S_2+\frac{1}{2}\log(L) $ for the bosonic Moore-Read state at cylinder perimeter $L=10$, plotted as a function of the charge $q$, for all three topological sectors, and for both fluxes $\Phi = 0$ (open markers) and $\Phi = 1/2$ (closed markers). The data for the vacuum and $\psi$ sectors is plotted with blue and green colorations that correspond to even and odd fermionic parity, respectively. This is done for both the MPS spectrum data and the fit synthetic entanglement spectrum data. For a particular flux, fermionic parity alternates between the vacuum and $\psi$ sectors with $q$.}
	\label{fig:MooreRead1S2qsubtractedbylihaldaneL10fitcombo}
\end{figure}

\begin{figure}
	\centering
	\includegraphics[scale=0.6]{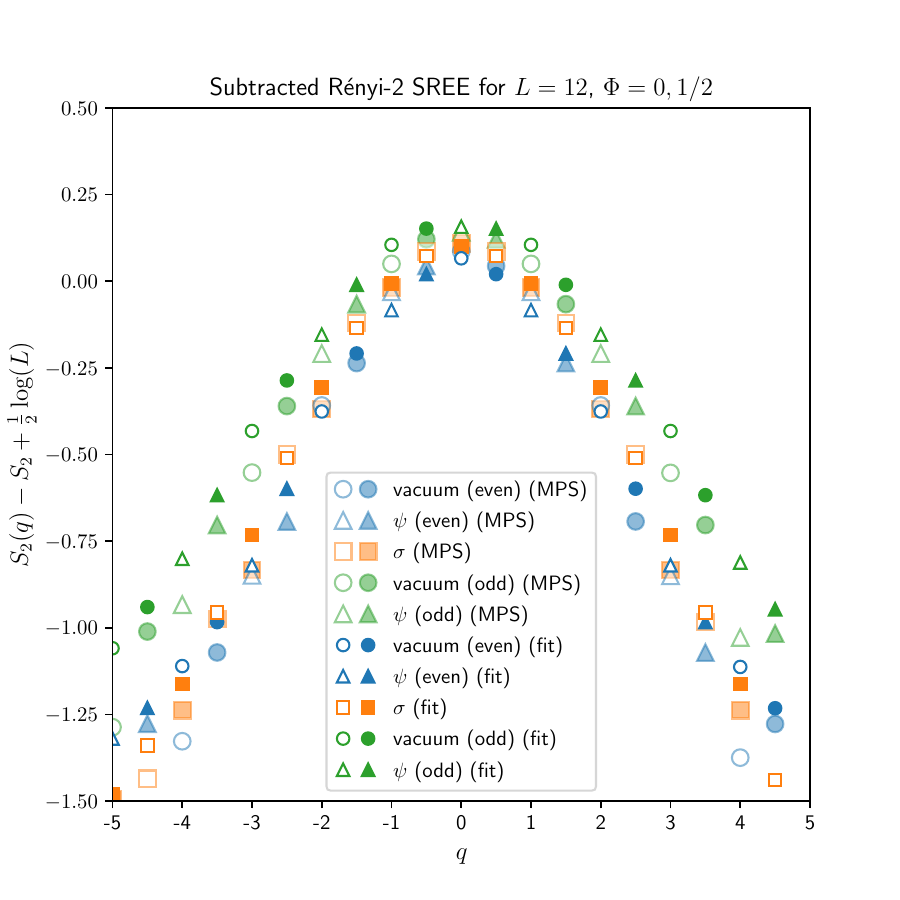}
	\caption{The subtracted symmetry-resolved second R\'{e}nyi entanglement entropy $S_2(q) -S_2+\frac{1}{2}\log(L) $ for the bosonic Moore-Read state at cylinder perimeter $L=12$, plotted as a function of the charge $q$, for all three topological sectors, and for both fluxes $\Phi = 0$ (open markers) and $\Phi = 1/2$ (closed markers). The data for the vacuum and $\psi$ sectors is plotted with blue and green colorations that correspond to even and odd fermionic parity, respectively. This is done for both the MPS spectrum data and the fit synthetic entanglement spectrum data. For a particular flux, fermionic parity alternates between the vacuum and $\psi$ sectors with $q$.}
	\label{fig:MooreRead1S2qsubtractedbylihaldaneL12fitcombo}
\end{figure}

Next, we calculate the subtracted symmetry-resolved entanglement entropy for the synthetic spectrum. For the second R\'{e}nyi entropy, this gives the results in Figs.~\ref{fig:MooreRead1S2qsubtractedbylihaldaneL10fitcombo} and \ref{fig:MooreRead1S2qsubtractedbylihaldaneL12fitcombo} at $L = 10$ and $L = 12$, respectively. 
In these figures, the comparable MPS data results are also shown, a comparison which reveals that the SREE from the synthetic ES does manage to capture most of the essential features, especially for small $|q|$.

\begin{figure*}
    \centering
    \subfloat[\label{fig:MooreRead1AllFullFCSL8fit}]{%
      \includegraphics[width=0.6666\columnwidth]{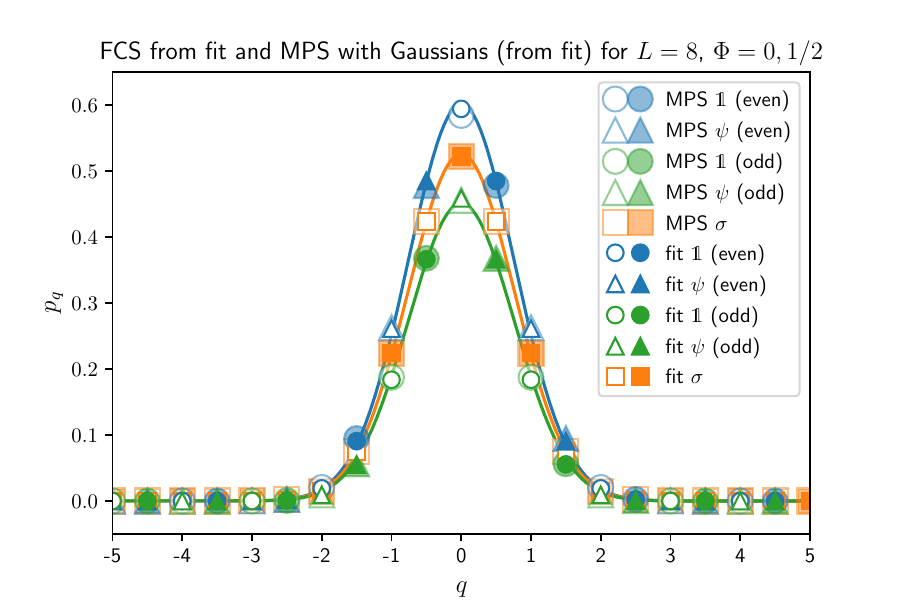}%
    }\hspace*{\fill}%
    \subfloat[\label{fig:MooreRead1AllFullFCSL10fit}]{%
      \includegraphics[width=0.6666\columnwidth]{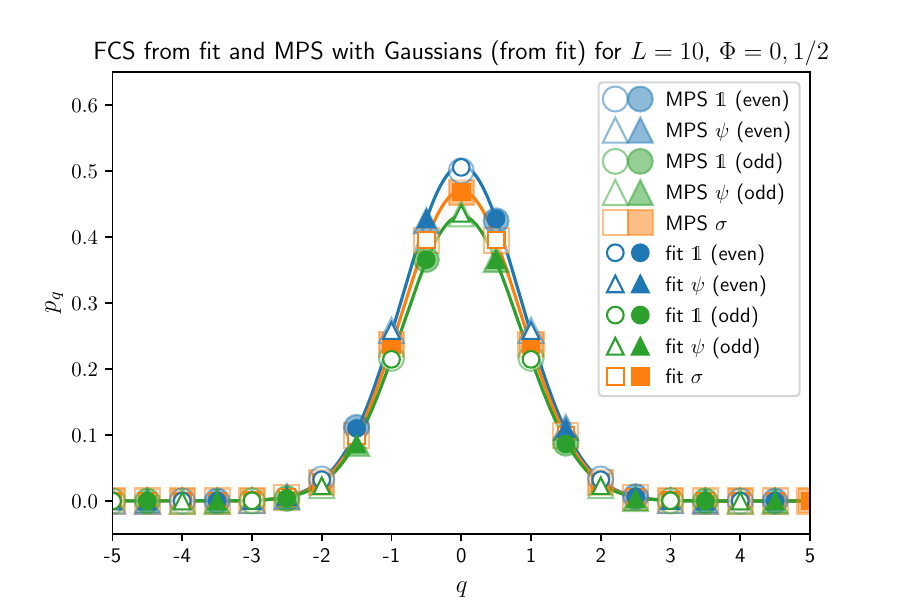}%
    }\hspace*{\fill}%
    \subfloat[\label{fig:MooreRead1AllFullFCSL12fit}]{%
      \includegraphics[width=0.6666\columnwidth]{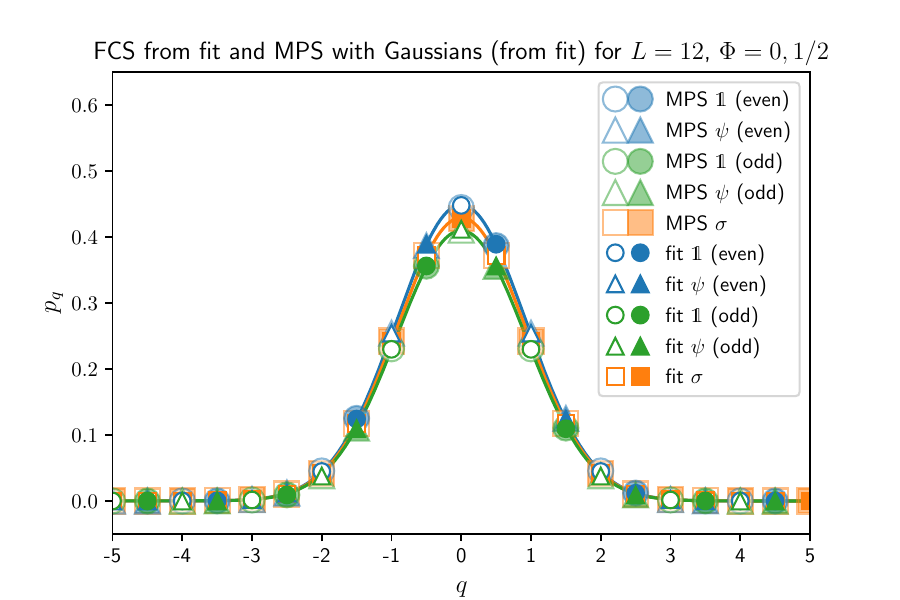}%
    }
    \caption{The FCS for the bosonic Moore-Read state plotted as a function $p_{q}$ of the charge $q$, for cylinder perimeters $L = 8$ (a), $L = 10$ (b), and $L = 12$ (c). The orange, square plot markers indicate the FCS from the non-Abelian $\sigma$ topological sector, while the circular and triangular plot markers indicate the Abelian topological sectors. The blue color corresponds to even fermionic parity, while the green color corresponds to odd fermionic parity; whether or not the data comes from the MPS entanglement spectrum or the fit synthetic entanglement spectrum is indicated by the size and shading of each data point. A clear overlap of the MPS and fit data is visible. We can also fit Gaussians to the FCS from the fit for the $\sigma$ sector and the even and odd parities over a range of cylinder perimeters $L$, from which we can then extract values for the bosonic and fermionic velocities. The realizations of these Gaussians for $L = 12$ are plotted here in the respective colors.}
    \label{fig:three_graphs_FCSfit}
\end{figure*}

We also again consider the FCS, now from the synthetic ES, shown in Figs.~\ref{fig:MooreRead1AllFullFCSL8fit}-\ref{fig:MooreRead1AllFullFCSL12fit} for cylinder perimeters $L = 8$, $10$, and $12$. In these figures, the MPS results are also shown for comparison, but the Gaussian fits are performed to the FCS data from the synthetic entanglement spectrum across the system sizes $L = 8,\ldots,12$. A clear overlap of the FCS from the two datasets can be seen. The fits are parametrized by what would be, in the Li-Haldane leading order, the bosonic and fermionic velocities $v_b$ and $v_f$, so in this way we also obtain estimates $v_{b,\textrm{FCS}} \approx 2.21$ and $v_{f,\textrm{FCS}} \approx 1.37$, which may be directly compared with the similar estimates from the FCS of the MPS data in Section \ref{sec:mpsresults}, which were $v_{b,\textrm{FCS}} \approx 2.19$ and $v_{f,\textrm{FCS}} \approx 1.34$.

\begin{figure}
	\centering
	\includegraphics[scale=0.6]{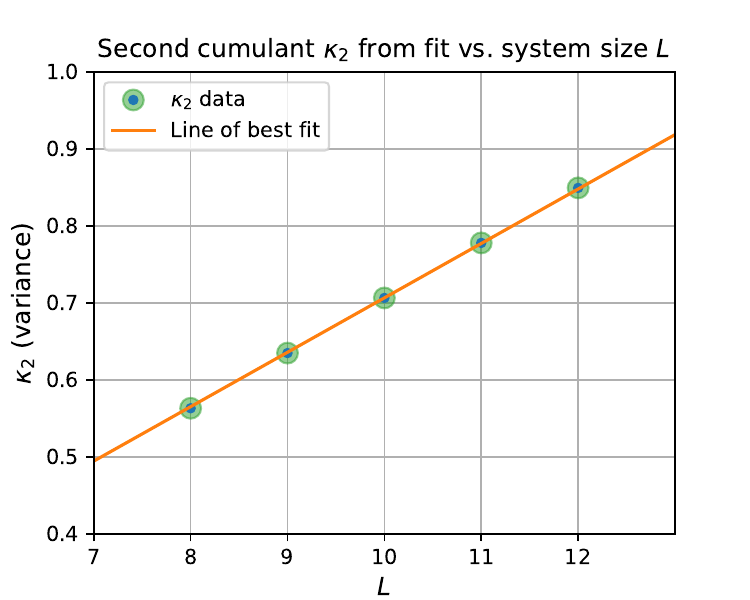}
	\caption{The second cumulant (variance) $\kappa_2$ of the FCS of the synthetic fit to the bosonic Moore-Read state, as seen in Figs.~\ref{fig:MooreRead1AllFullFCSL8fit}-\ref{fig:MooreRead1AllFullFCSL12fit}, as a function of cylinder perimeter $L$, for $L = 8,\ldots,12$. The line of best fit, which has a fixed $\kappa_2$-intercept at zero, is fit to the data, and has slope approximately 0.0707, corresponding to a $v_b$ estimate of around 2.25.}
	\label{fig:MooreRead1fcskappa2fit}
\end{figure}

Yet another estimate of the bosonic velocity $v_b$ may be obtained from directly computing the second cumulant/variance $\kappa_2 =\sigma^2$ of the FCS and using Eq.~\eqref{eq:sigma2L} to extract $v_{b,\kappa_2}$ from the slope of a line of best fit of the $\kappa_2$ vs.~$L$ graph. This is shown for the second cumulant data of the synthetic ES in Fig.~\ref{fig:MooreRead1fcskappa2fit}, where we find a Li-Haldane leading order estimate of $v_{b,\kappa_2} \approx 2.25$, very close indeed to the $v_{b,\kappa_2} \approx 2.24$ estimate from the MPS data. 

Like the parity imbalance estimates $v_{b,\textrm{parity}}$ and $v_{f,\textrm{parity}}$ discussed in Section \ref{sec:ESfitresults}, there is also a substantial discrepancy between the $v_{b,\textrm{FCS}}$ and $v_{f,\textrm{FCS}}$, and $v_{b,\kappa_2}$ estimates, and the synthetic ES fit parameters $v_b$ and $v_f$ of Eq.~\eqref{eq:syntheticvbvf}. Again, this can be mostly accounted for by contributions to the parity imbalance from the contributions of the integrals of higher-order operators, those with some bosonic content contributing to $v_{b,\textrm{FCS}}$ and $v_{b,\kappa_2}$, and those with fermionic content to $v_{f,\textrm{FCS}}$. The synthetic ES approach is able to go somewhat further than these Li-Haldane leading order estimates in determining the actual $v_b$ and $v_f$ values, by separating out some of the higher-order contributions. We combine all of the various estimates for $v_b$ and $v_f$ from both the MPS and synthetic ES data in Table \ref{table:vbvfestimates}.

\begin{table}[]
    \centering
    \begin{tabular}{c|c||c|c}
    \multicolumn{2}{c||}{Quantity fit} & $v_b$ & $v_f$ \\
    \hline
    \hline
    parity & MPS & 2.11 & 1.40 \\
    \cline{2-4}
    $\Phi = 0$ & synthetic & 2.17 & 1.42 \\
    \hline
    parity & MPS & --- & 1.40\\
    \cline{2-4}
    $\Phi = 1/2$ & synthetic & --- & 1.42\\
    \hline
    \multirow{2}{*}{FCS} & MPS & 2.19 & 1.34 \\
    \cline{2-4}
     & synthetic & 2.21 & 1.37  \\
    \hline
    \multirow{2}{*}{$\kappa_2$} & MPS & 2.24 & --- \\
    \cline{2-4}
     & synthetic & 2.25 & --- \\
    \hline \hline
    ES & synthetic & 1.82 & 0.774 
    \end{tabular}
    \caption{In this table, we collect the various estimates of the bosonic velocity $v_b$ and fermionic velocity $v_f$ of the leading order $H_A$ of Eq.~\eqref{eq:asymmetriclihaldanesubbed} that come from fitting various quantities in both the MPS and synthetic ES numerical data. Notably, the final row's estimates, which come directly from the fit of the synthetic entanglement spectrum of Eq.~\eqref{eq:syntheticvbvf}, are distinct from the others due to the fact that they also account for the operators of conformal dimension $\Delta_i = 4$ in Table \ref{table:integerevenchargeoperators}.}
    \label{table:vbvfestimates}
\end{table}

\end{document}